\documentclass[%
 reprint,
superscriptaddress,
 amsmath,amssymb,
 aps,
]{revtex4-2}

\usepackage{graphicx}% Include figure files
\usepackage{tikz}
\usepackage{dcolumn}% Align table columns on decimal point
\usepackage{bm}% bold math
\allowdisplaybreaks[2]
\usepackage[colorlinks,linkcolor=blue, urlcolor=blue, anchorcolor=blue, citecolor=blue]{hyperref}% add hypertext capabilities
\begin{document}

\preprint{APS/123-QED}

\title{Generating Bell states and Werner states of two qubits via optical field}% Force line breaks with \\
%\thanks{A footnote to the article title}%

\author{Dengkui Jiang}
\affiliation{Key Laboratory of Low-Dimensional Quantum Structures and Quantum Control of Ministry of Education, Department of Physics
	and Synergetic Innovation Center for Quantum Effects and Applications, Hunan Normal University, Changsha 410081, China}

\author{Cuilu Zhai}
\affiliation{Department of Maths and Physics, Hunan Institute of Engineering, Xiangtan 411104, China}

\author{Yaju Song}
\affiliation{College of Physics and Electronic Engineering, Hengyang Normal University, Hengyang 421002, China}

\author{Zhaohui Peng}
\affiliation{Hunan Provincial Key Laboratory of Intelligent Sensors and Advanced Sensor Materials, and Department of
	Physics, Hunan University
	of Science and Technology,
	Xiangtan 411201, China}

\author{Jibing Yuan}
\affiliation{College of Physics and Electronic Engineering, Hengyang Normal University, Hengyang 421002, China}

\author{Shiqing Tang}
\email{sqtang@hynu.edu.cn}
\affiliation{College of Physics and Electronic Engineering, Hengyang Normal University, Hengyang 421002, China}

\author{Wangjun Lu}
\email{wjlu1227@zju.edu.cn}
\affiliation{Key Laboratory of Low-Dimensional Quantum Structures and Quantum Control of Ministry of Education, Department of Physics
	and Synergetic Innovation Center for Quantum Effects and Applications, Hunan Normal University, Changsha 410081, China}
\affiliation{Department of Maths and Physics, Hunan Institute of Engineering, Xiangtan 411104, China}
\affiliation{College of Physics and Electronic Engineering, Hengyang Normal University, Hengyang 421002, China}

%\collaboration{CLEO Collaboration}%\noaffiliation

\date{\today}% It is always \today, today,
             %  but any date may be explicitly specified

\begin{abstract}
In this paper, we investigate how the evolution of the states of two qubits initially in a direct product state can be controlled by the optical field in a Tavis-Cummings (TC) model. For the two qubits initially in the direct product state, we find that their matrix elements at any moment can be modulated by the coefficients of the optical field initial states in the number state space. We propose a method for preparing an \textit{X}-type state of two qubits. Subsequently, for descriptive convenience, we divide the Bell states of the two qubits into two kinds in the paper. When both qubits are initially in the ground state, we find that the two qubits can be controlled to produce the first type of Bell state by the superposition state optical field that is initially in the next-nearest-neighbor number state and that the production of any of the first type of Bell states can be controlled by controlling the phase between the two next-nearest-neighbor number states. When one of the two qubits is in the ground state, and the other is in the excited state, we can control the two qubits to produce the second type of Bell state by the single-photon number state optical field. Finally, we study the generation of Werner states by controlling two qubits initially, both in the ground state, using an optical field.
\end{abstract}

%\keywords{Suggested keywords}%Use showkeys class option if keyword
                              %display desired
\maketitle

%\tableofcontents

\section{introduction}
Quantum entanglement is one of the core features of quantum mechanics and has an extremely important impact on the field of quantum information science \cite{PhysRev.47.777,PhysRev.48.696, RevModPhys.81.865, peres1997quantum}. Quantum entanglement is widely used in areas such as quantum computing \cite{nielsen2010quantum, ladd2010quantum, berman1998introduction,preskill1998reliable,PhysRevX.10.041038,steane1998quantum}, quantum communication \cite{gisin2007quantum, cozzolino2019high,chen2021integrated,yuan2010entangled,nauerth2013air,pirandola2017fundamental,long2007quantum,duan2001long, dai2008teleportation,peng2013construction,RevModPhys.94.035001}, and quantum cryptography \cite{pirandola2020advances,bennett1992quantum, bennett1992experimental, RevModPhys.74.145, RevModPhys.94.025008, bernstein2009introduction,PhysRevLett.68.3121,bernstein2017post,groblacher2006experimental,hughes1995quantum,PhysRevA.59.3301,PhysRevLett.75.1239,PhysRevLett.85.1330,PhysRevA.61.062308,yin2020entanglement}. Among the states of quantum entanglement, the simplest and most popular form of maximum quantum entanglement is the Bell state \cite{nielsen2010quantum,RevModPhys.86.419,sych2009complete,gisin1998bell}, also known as the EPR pair \cite{PhysRev.47.777,kumar2008quantum}. There are four Bell states, each representing a maximally entangled pair of qubits. These states are $|\Phi_{\pm}\rangle=(|gg\rangle\pm|ee\rangle)/\sqrt{2}$ and $|\Psi\rangle=(|ge\rangle\pm|eg\rangle)/\sqrt{2}$. Here, $|g\rangle$ and $|e\rangle$ denote the ground and excited states of the two-level system, respectively. Bell states are used in many quantum computing and quantum communication protocols \cite{sisodia2017experimental,walther2005experimental,PhysRevA.73.022330,engel2005fermionic,PhysRevA.76.012335,barz2013experimental,PhysRevLett.95.020403,whitaker2012new,o2007optical,PhysRevA.82.032318,PhysRevA.60.157}, such as quantum teleportation \cite{bouwmeester1997experimental,riebe2004deterministic,pirandola2015advances,zeilinger2000quantum,metcalf2014quantum,PhysRevA.93.062305}, superdense coding \cite{barreiro2008beating,PhysRevA.77.032321,PhysRevA.65.022304,PhysRevLett.118.050501,PhysRevA.75.060305,hu2018beating}, and various forms of quantum cryptography \cite{bennett1992quantum, PhysRevLett.84.4733,PhysRevA.73.050302}. Moreover, Werner states proposed by \textit{Werner} are also a very important quantum state \cite{PhysRevA.40.4277}. Unlike pure quantum states, Werner states represent a class of mixed states that can be either entangled or separable, depending on certain parameters. Werner states are used in quantum information theory to explore the boundaries between classical and quantum correlations and to study the robustness of entanglement under various quantum operations and in the presence of decoherence \cite{PhysRevA.62.044302,PhysRevLett.84.4236,PhysRevA.90.012114}. 

Due to the very important applications of Bell and Werner states, how to prepare these quantum states in real quantum systems is a very crucial step in the applications of quantum information science. It is encouraging to note that the preparation of these states has been realized in various types of quantum systems. For example, \textit{Kim et al.} realized the preparation of optical Bell states using Bell state measurement in an optical system \cite{kim2018informationally}, \textit{Zhang et al. } investigated the evolution of two qubits, both in the ground state, to Bell states in a superconducting waveguide quantum electrodynamics system driven by an appropriate microwave pulse \cite{PhysRevApplied.20.044014}, \textit{Macr\`{i} et al.} proposed the preparation of qubits in the ultrastrong-coupling circuit QED to realize Bell state preparation \cite{PhysRevA.98.062327}, and \textit{Clark et al.} realized Bell state preparation in trapped-ion systems \cite{PhysRevLett.127.130505}. In addition, the realization of Bell state preparation in other quantum systems has been extensively studied \cite{PhysRevA.78.054301,PhysRevA.69.042303,PhysRevA.105.022624,PhysRevLett.118.030503,PhysRevLett.110.160502,PhysRevResearch.3.043031,PhysRevLett.111.100406,PhysRevA.63.062307,PhysRevLett.124.053602,PhysRevLett.130.053601,PhysRevLett.128.080503}. Meanwhile, dissipative systems can also realize Bell state preparation \cite{PhysRevB.106.L180406,PhysRevApplied.18.014051}. Werner states can be prepared through spontaneous parametric down-conversion or a Universal Source of Entanglement method \cite{PhysRevA.66.062315,PhysRevLett.92.177901}.

With the development of experimental techniques in cavity quantum electrodynamics, it has become possible to control atom-light interactions precisely \cite{haroche1989cavity,walther2006cavity,mabuchi2002cavity,RevModPhys.85.1083,RevModPhys.73.565,mirhosseini2019cavity}. The generation of maximally entangled states has been realized in a series of TC models under different conditions \cite{PhysRevLett.85.2392,ficek2002entangled, PhysRevLett.110.083603,PhysRevA.68.052312,casagrande2008generation,PhysRevA.108.023728,PhysRevA.104.063701}, such as the TC model under large detuning conditions \cite{PhysRevLett.85.2392}, the nonlinear two-atom TC model \cite{PhysRevA.104.063701}, the TC model with driven dissipation \cite{casagrande2008generation}, and so on. Unlike the above studies, in this paper, we investigate how to control the initial state of the light field to control the two qubits in the direct product state to produce the maximally entangled state in the TC model under resonance conditions by an exact solution without any approximation. We propose a method to control the generation of Bell and Werner states from two two-level atoms in a cavity quantum electrodynamic system using an optical field. By analytically solving the  TC model under resonance conditions, we find that the density matrices of the two two-level atoms can be modulated by the expanding coefficients of the optical field initial states on the number-state basis vectors. Based on this result, we first propose a method to prepare \textit{X}-type quantum states. Then, we categorize the maximally entangled states of two qubits into two kinds, $|ee\rangle$ and $|gg\rangle$ superposition, and $|eg\rangle$ and $|ge\rangle$ superposition, the former of which we refer to as the first type of Bell states, and the latter of which we refer to as the second type of Bell states, and then we propose a method to prepare these two types of Bell states by controlling the optical field initial state. Finally, we propose methods to prepare Werner states. Our method is experimentally realizable for the preparation of these states.

\section{Model and solution}

\begin{figure}[t]
\centering
\includegraphics[width=6cm,height=4cm]{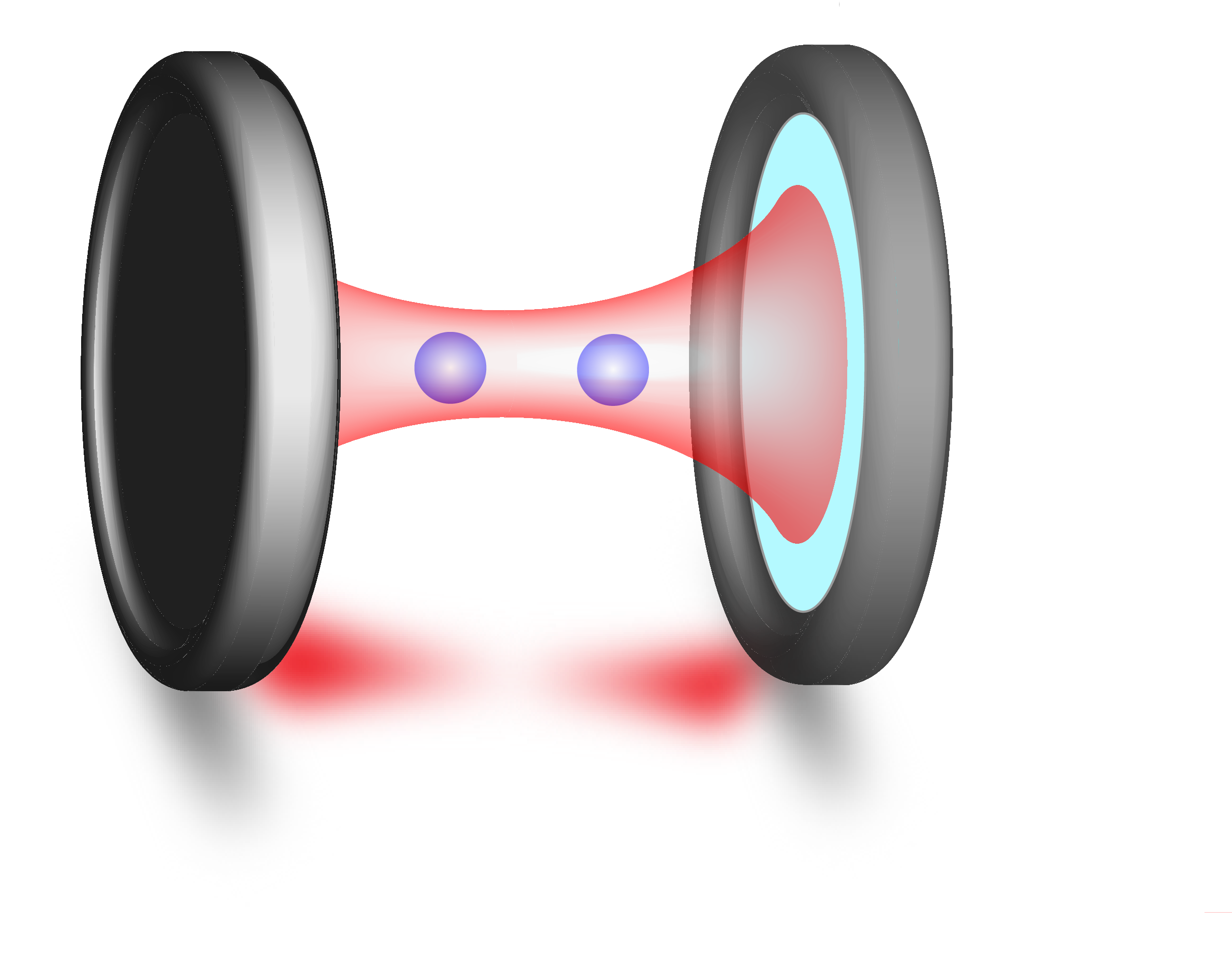}	
\caption{\label{fig1}Schematic diagram of the two qubits TC model. The two green balls denote two qubits. The interaction exists only between the qubit and the optical field; there is no interaction between the two qubits.}
\end{figure}

In this article, we investigate the control of the quantum state of the qubits system by the optical field in a single-mode optical field and two identical  qubit-interacting systems (as shown in Fig.~\ref{fig1}). We assume that there is no interaction between qubits and photons, and the interaction exists only between the qubit system and the optical field, and this interaction is a dipole interaction. Under near-resonance and relatively weak coupling conditions, the Hamiltonian of the whole system is described by the following Tavis-Cummings (TC) model Hamiltonian ($\hbar=1$).
\begin{equation}
\hat{H}_{TC}=\text{\ensuremath{\omega_{a}}}\hat{a}^{\dagger}\hat{a}+\frac{1}{2}\omega_{0}\sum_{i=1}^{N}\hat{\sigma}_{z}^{i}+g\sum_{i=1}^{N}(\hat{a}^{\dagger}\hat{\sigma}_{-}^{i}+\hat{\sigma}_{+}^{i}\hat{a}),  \label{Eq1}
\end{equation}
where $\omega_{a}$ is the eigenfrequency of the optical cavity, $\hat{a}^{\dagger}$ and $\hat{a}$ are the usual creation and annihilation operators of bosons, respectively, which satisfy the commutation relation $[\hat{a}, \hat{a}^{\dagger}]=1$, $[\hat{a}, \hat{a}]=[\hat{a}^{\dagger}, \hat{a}^{\dagger}]=0$. The transition frequency between two energy levels of a single qubit in $N$ identical qubits is $\omega_{0}$. $g$ is the coupling strength between the single-mode optical field and a single qubit. $\hat{\sigma}_{\pm}^{i}=\frac{1}{2}(\hat{\sigma}_{x}^{i}\pm i\hat{\sigma}_{y}^{i})$,  $\hat{\sigma}_{x}^{i}$, $\hat{\sigma}_{y}^{i}$, and $\hat{\sigma}_{z}^{i}$ are the usual Pauli operators of the $i$th qubit.

We first study the case where a single-mode optical field interacts with two identical qubits ($N=2$). Under the resonant condition $\omega_{a}=\omega_{0}$ and in the interaction picture, the unitary evolution operator of the system at time $t$ is denoted as \cite{fujii2004explicit}
\begin{eqnarray}
\ensuremath{\hat{U}(t)=\exp(-i\hat{H}_{int}t)=\left[\begin{array}{cccc}
	\hat{U}_{11} & \hat{U}_{12} & \hat{U}_{13} & \hat{U}_{14}\\
	\hat{U}_{21} & \hat{U}_{22} & \hat{U}_{23} & \hat{U}_{24}\\
	\hat{U}_{31} & \hat{U}_{32} & \hat{U}_{33} & \hat{U}_{34}\\
	\hat{U}_{41} & \hat{U}_{42} & \hat{U}_{43} & \hat{U}_{44}
	\end{array}\right]}, \label{Eq2}
\end{eqnarray}
where $\hat{H}_{int}=g\sum_{i=1}^{2}(\hat{a}^{\dagger}\hat{\sigma}_{-}^{i}+\hat{\sigma}_{+}^{i}\hat{a})$ and the matrix elements in Eq.~(\ref{Eq2}) are as follows
\begin{eqnarray}
\hat{U}_{11}&=&1+2\frac{\hat{A}(\hat{n}+1)-1}{\hat{C}(\hat{n}+1)}(\hat{n}+1), \label{Eq3}\\	\hat{U}_{44}&=&1+2\frac{\hat{A}(\hat{n}-1)-1}{\hat{C}(\hat{n}-1)}\hat{n}, \label{Eq4} \\
\hat{U}_{22}&=&\hat{U}_{33}=\frac{\hat{A}(\hat{n})+1}{2}, \label{Eq5}\\	\hat{U}_{23}&=&\hat{U}_{32}=\frac{\hat{A}(\hat{n})-1}{2}, \label{Eq6} \\
\hat{U}_{14}&=&2\frac{\hat{A}(\hat{n}+1)-1}{\hat{C}(\hat{n}+1)}\hat{a}^{2},  \label{Eq7}\\    	\hat{U}_{41}&=&2\frac{\hat{A}(\hat{n}-1)-1}{\hat{C}(\hat{n}-1)}\hat{a}^{\dagger2}, \label{Eq8}\\
\hat{U}_{12}&=&\hat{U}_{13}=-i\frac{\hat{B}(\hat{n}+1)}{\sqrt{\hat{C}(\hat{n}+1)}}\hat{a}, \label{Eq9}\\	\hat{U}_{21}&=&\hat{U}_{31}=-i\frac{\hat{B}(\hat{n})}{\sqrt{\hat{C}(\hat{n})}}\hat{a}^{\dagger}, \label{Eq10} \\
\hat{U}_{42}&=&\hat{U}_{43}=-i\frac{\hat{B}(\hat{n}-1)}{\sqrt{\hat{C}(\hat{n}-1)}}\hat{a}^{\dagger},	\label{Eq11}\\
\hat{U}_{24}&=&\hat{U}_{34}=-i\frac{\hat{B}(\hat{n})}{\sqrt{\hat{C}(\hat{n})}}\hat{a}, \label{Eq12} \\
\hat{A}(\hat{n})&=&\cos(gt\sqrt{\hat{C}(\hat{n})}),	\hat{B}(\hat{n})=\sin(gt\sqrt{\hat{C}(\hat{n})}), \nonumber \\
\hat{C}(\hat{n})&=&2(2\hat{n}+1). \nonumber 
\end{eqnarray}
In all the above equations, $\hat{n}$ denotes the number operator of the optical field.

In the following, we focus on the variation of the two identical qubits' quantum states over time for different optical field initial states. First, we prepare two identical qubits in the ground state $|g\rangle$ and the optical field in an arbitrary state $|\phi_{a}(0)\rangle$, then the initial state of the total system is
\begin{eqnarray}
|\Psi(0)\rangle=|g\rangle\otimes|g\rangle\otimes|\phi_{a}(0)\rangle=\left[\begin{array}{c}
0\\
0\\
0\\
|\phi_{a}(0)\rangle
\end{array}\right].  \label{Eq13}
\end{eqnarray}
The state of the total system at time $t$ is
\begin{eqnarray}
|\Psi(t)\rangle&=&\hat{U}(t)|\Psi(0)\rangle \nonumber\\
&=&\left[\begin{array}{cccc}
\hat{U}_{11} & \hat{U}_{12} & \hat{U}_{13} & \hat{U}_{14}\\
\hat{U}_{21} & \hat{U}_{22} & \hat{U}_{23} & \hat{U}_{24}\\
\hat{U}_{31} & \hat{U}_{32} & \hat{U}_{33} & \hat{U}_{34}\\
\hat{U}_{41} & \hat{U}_{42} & \hat{U}_{43} & \hat{U}_{44}
\end{array}\right]\left[\begin{array}{c}
0\\
0\\
0\\
|\phi_{a}(0)\rangle
\end{array}\right] \nonumber\\
&=&\left[\begin{array}{c}
\hat{U}_{14}|\phi_{a}(0)\rangle\\
\hat{U}_{24}|\phi_{a}(0)\rangle\\
\hat{U}_{34}|\phi_{a}(0)\rangle\\
\hat{U}_{44}|\phi_{a}(0)\rangle
\end{array}\right]. \label{Eq14}
\end{eqnarray}
Meanwhile, by tracing off the states of the optical field, we can obtain the density matrix of the two qubits at time $t$ as follows
\begin{eqnarray}
\hat{\rho}_{q}(t)&=&\textbf{Tr}_{a}[|\Psi(t)\rangle\langle\Psi(t)|] \nonumber\\
&=&\sum_{n=0}^{\infty}\langle n|\Psi(t)\rangle\langle\Psi(t)|n\rangle \nonumber \\
&=&\left[\begin{array}{cccc}
v_{+} & h_{+}^{*} & h_{+}^{*} & \mu^{*}\\
h_{+} & w & p & h_{-}^{*}\\
h_{+} & p & w & h_{-}^{*}\\
\mu & h_{-} & h_{-} & v_{-}
\end{array}\right],  \label{Eq15}
\end{eqnarray}
where $\textbf{Tr}_{a}[\cdot]$ denotes tracing off the optical field state of the optical-qubits system, $|n\rangle$ denotes the number state of the optical field, and
\begin{eqnarray}
v_{+}&=&\sum_{n=0}^{\infty}\left|c_{n+2}\right|^{2}4(n+2)(n+1)\left[\frac{A(n+1)-1}{C(n+1)}\right]^{2},  \label{Eq16}\\
h_{+}&=&\sum_{n=0}^{\infty}c_{n+1}c_{n+2}^{*}(-2i(n+1))\sqrt{(n+2)}\frac{B(n)}{\sqrt{C(n)}} \nonumber\\
&&\times\frac{A(n+1)-1}{C(n+1)} , \label{Eq17}\\
h_{-}&=&\sum_{n=0}^{\infty}c_{n}c_{n+1}^{*}(i\sqrt{n+1}\frac{B(n)}{\sqrt{C(n)}})\Big(2\frac{A(n-1)-1}{C(n-1)}n \nonumber\\
&&+1\Big),  \label{Eq18}\\
\mu&=&\sum_{n=0}^{\infty}c_{n}c_{n+2}^{*}2\sqrt{(n+2)(n+1)}\frac{A(n+1)-1}{C(n+1)}\nonumber\\
&&\times\left(1+2\frac{A(n-1)-1}{C(n-1)}n\right), \label{Eq19}\\
w&=&p=\sum_{n=0}^{\infty}\left|c_{n+1}\right|^{2}(n+1)\frac{B^{2}(n)}{C(n)}, \label{Eq20}\\
v_{-}&=&\sum_{n=0}^{\infty}\left|c_{n}\right|^{2}\left(1+2\frac{A(n-1)-1}{C(n-1)}n\right)^{2}, \label{Eq21}\\
A(n)&=&\cos(gt\sqrt{C(n)}), B(n)=\sin(gt\sqrt{C(n)}),\nonumber \\
C(n)&=&2(2n+1).\nonumber
\end{eqnarray}
where $c_{n}=\langle n|\phi_{a}(0)\rangle$, $c_{n}^{*}$ denotes the complex conjugate of $c_{n}$, and the detailed calculations from Eq.~(\ref{Eq16}) to Eq.~(\ref{Eq21}) are shown in Appendix \ref{Appendix A}. From Eq.~(\ref{Eq16}) to Eq.~(\ref{Eq21}), we can see that the end states of the two qubits can be controlled by choosing optical field state with different probability distributions on the number state space. For example, if the end state of the two qubits is an \textit{X}-type state, the initial state of the optical field can be chosen as follows
\begin{equation}
|\phi_{a}(0)\rangle=\sum_{n=0}^{\infty}c_{n}|n\rangle,  \label{Eq22}
\end{equation}
where $c_{n}=\langle n|\phi_{a}(0)\rangle$, and these coefficients satisfy $c_{n}c_{n+1}=0$. That is, at least one of the probability distributions of this initial state over adjacent number states is zero. Such an initial optical field can lead to the matrix element of the density matrix of the two qubits at the moment $t$ to satisfy $h_{+}=h_{-}=0$, i.e., the density matrix of the two qubits at an arbitrary moment under such an optical field is an \textit{X}-type density matrix as follows
\begin{eqnarray}
\hat{\rho}_{q}(t)=\left[\begin{array}{cccc}
v_{+} & 0 & 0 & \mu^{*}\\
0 & w & w & 0\\
0 & w & w & 0\\
\mu & 0 & 0 & v_{-}
\end{array}\right].  \label{Eq23}
\end{eqnarray}
It is particularly noteworthy that the superposition of any number states, as shown in Eq.~(\ref{Eq22}), can be experimentally realized  \cite{PhysRevLett.118.223604}. Since the two qubits are identical, we have $w=p$. In the following representation of the density matrix for the two identical qubits, we will consistently use the notation $w=p$. 

\section{Preparation of the first class of Bell states}

By analyzing the matrix elements of the density matrix of the two qubits at time $t$, we have already learned how to control the initial state of the optical field in the number state space to prepare an \textit{X}-type state of the two qubits. For example, we can choose the initial state of the optical field to be either an odd or an even coherent state. Of course, we can also choose other number-state superposition states, such as
\begin{equation}
|\phi_{a}(0)\rangle=c_{m}|m\rangle+c_{m+2}|m+2\rangle , \label{Eq24}
\end{equation}
where $|c_{m}|^{2}+|c_{m+2}|^{2}=1$. We refer to this state as a superposition of next-nearest-neighbor number states. In experiments, it is feasible to realize arbitrary number state superpositions \cite{PhysRevLett.118.223604}.

Next, we begin to study the generation of the maximum entangled state of two identical qubits by controlling the initial state of the optical field. In order to prepare a maximum entangled state of two qubits like 
\begin{equation}
|B_{1}\rangle=(|ee\rangle+e^{i\phi}|gg\rangle)/\sqrt{2},  \label{Eq25}
\end{equation}
where $\phi$ is the relative phase between $|ee\rangle$ and $|gg\rangle$. The density matrix of this maximal entangled state above is of the form
\begin{eqnarray}
\hat{\rho}_{m_{1}}&=&|B_{1}\rangle\langle B_{1}| \nonumber \\
&=&\left[\begin{array}{cccc}
0.5 & 0 & 0 & e^{-i\phi}/2\\
0 & 0 & 0 & 0\\
0 & 0 & 0 & 0\\
e^{i\phi}/2 & 0 & 0 & 0.5
\end{array}\right].  \label{Eq26}
\end{eqnarray}
In the following, we choose the initial state of the optical field as $|\phi_{a}(0)\rangle=c_{m}|m\rangle+c_{m+2}|m+2\rangle$, and from the above analysis, we can get $h_{+}=h_{-}=0$, and the other matrix elements of the two qubits density matrix are as follows
\begin{eqnarray}
v_{+}&=&\left|c_{m}\right|^{2}4m(m-1)\left[\frac{A(m-1)-1}{C(m-1)}\right]^{2}+\left|c_{m+2}\right|^{2} \nonumber \\
&&\times 4(m+2)(m+1)\left[\frac{A(m+1)-1}{C(m+1)}\right]^{2},  \label{Eq27}\\
w &=& p =|c_{m}|^{2}m\frac{B^{2}(m-1)}{C(m-1)} +|c_{m+2}|^{2}(m+2)\nonumber \\
&& \times \frac{B^{2}(m+1)}{C(m+1)},  \label{Eq28}\\  
\mu&=&c_{m}c_{m+2}^{*}2\sqrt{(m+2)(m+1)}\frac{A(m+1)-1}{C(m+1)}\nonumber\\
&&\times\left(1+2\frac{A(m-1)-1}{C(m-1)}m \right) ,  \label{Eq29}\\   
v_{-}&=&\left|c_{m}\right|^{2}\left(1+2m\frac{A(m-1)-1}{C(m-1)}\right)^{2}+\left|c_{m+2}\right|^{2} \nonumber \\
&&\left(1+2(m+2)\frac{A(m+1)-1}{C(m+1)}\right)^{2} . \label{Eq30}
\end{eqnarray}

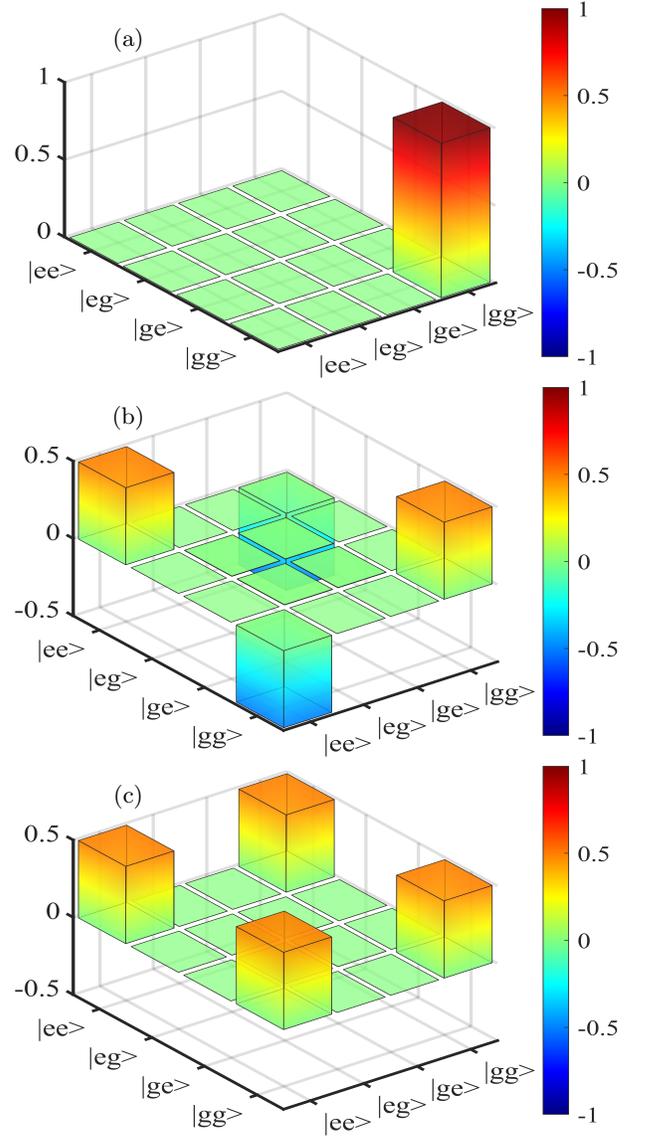
\begin{figure}[t]
\centering
\begin{tikzpicture}
\draw (0, 0) node[inner sep=0] {\includegraphics[width=8cm,height=5cm]{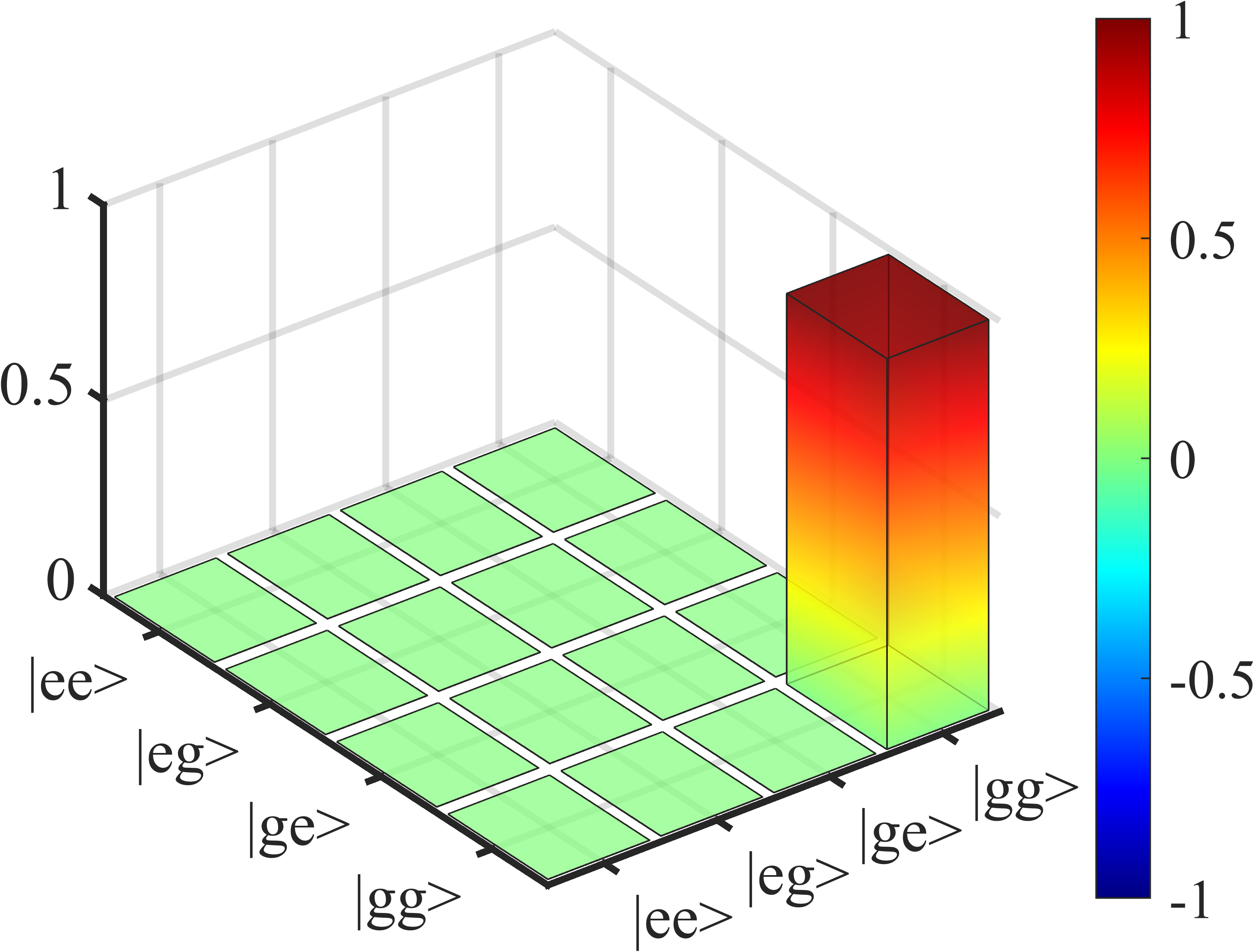}};
\draw (-2.5, 2) node {(a)};
\end{tikzpicture}
\begin{tikzpicture}
\draw (0, 0) node[inner sep=0] {\includegraphics[width=8cm,height=5cm]{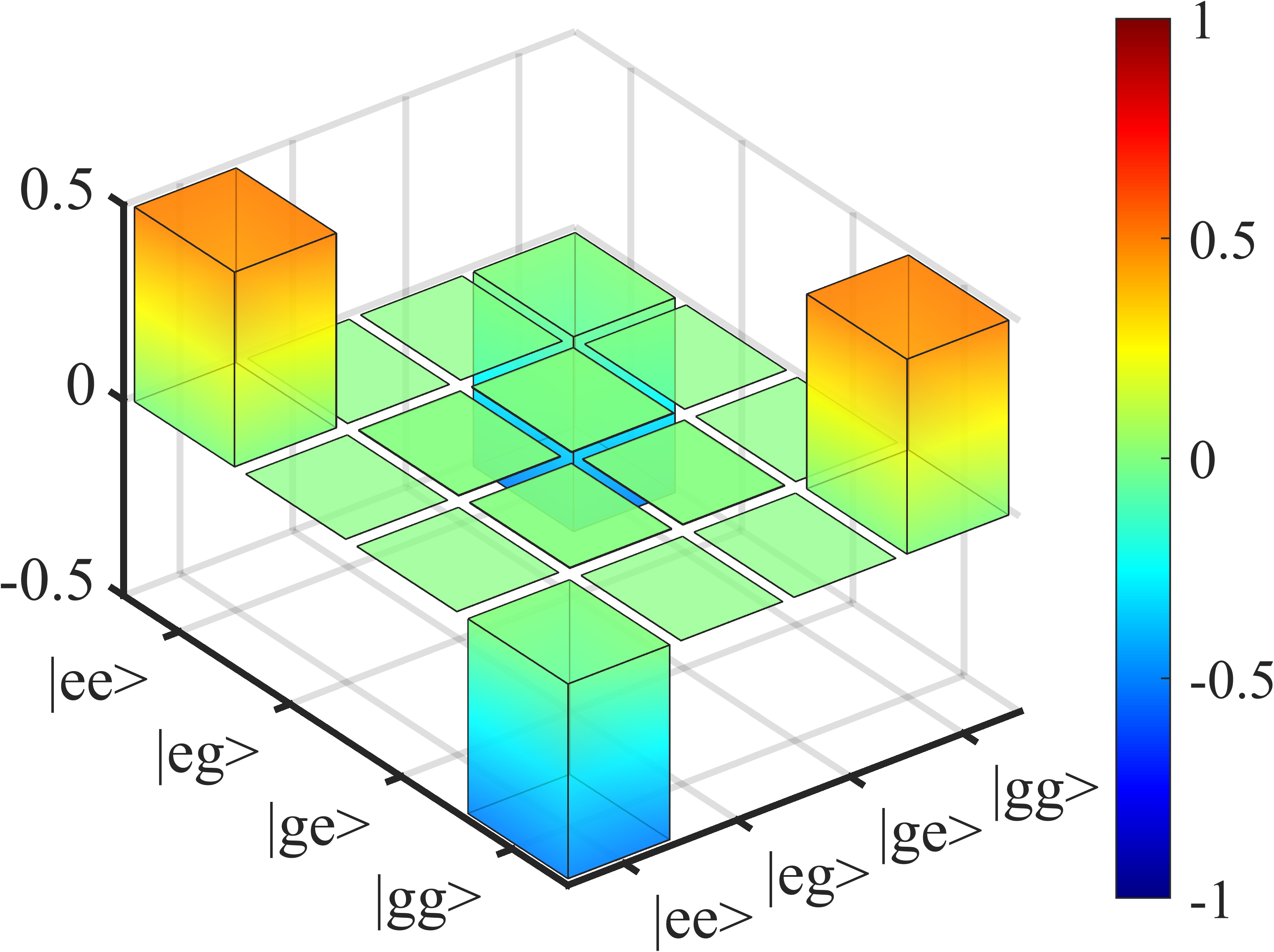}};
\draw (-2.5, 2) node {(b)};
\end{tikzpicture}
\begin{tikzpicture}
\draw (0, 0) node[inner sep=0] {\includegraphics[width=8cm,height=5cm]{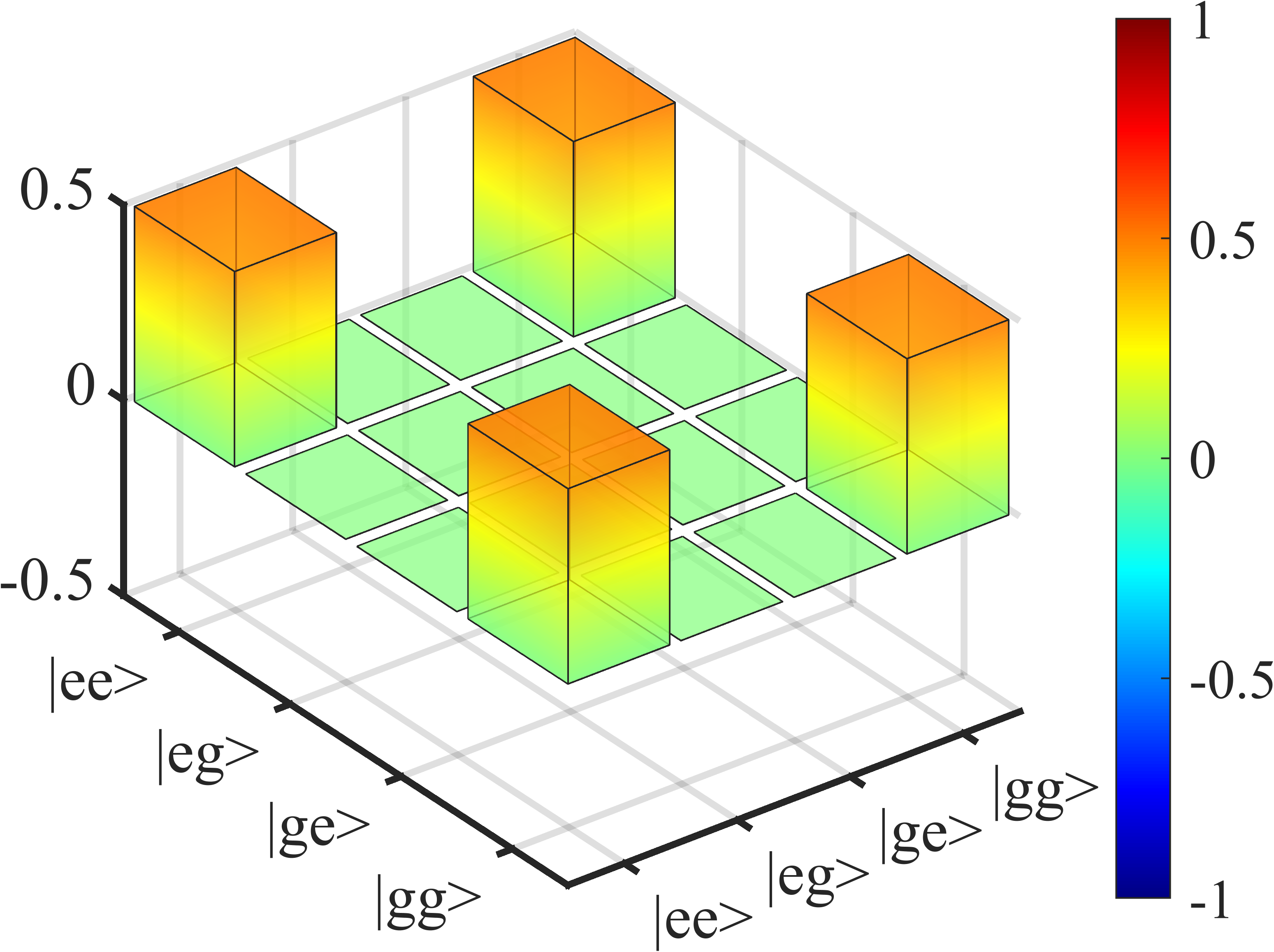}};
\draw (-2.5, 2) node {(c)};
\end{tikzpicture}
\caption{\label{fig2} When the initial state of the two qubits is $|gg\rangle$, (a) denotes the matrix elemental image of the initial density matrix of the two qubits, and (b) and (c) denote the matrix elemental images of the two qubits when they reach the maximally entangled state for the first time under the control of the optical field in the initial state of $|\phi_{a}(0)\rangle=(|30\rangle+ |32\rangle)/\sqrt{2}$ and the optical field in the initial state of $|\phi_{a}(0)\rangle=(|40\rangle-|42\rangle)\sqrt{2}$, respectively.}
\end{figure}

If we need to obtain a two-qubits maximally entangled pure state as shown in Eq.~(\ref{Eq25}), then $w = p = 0$, i.e.,  
\begin{equation}
B(m-1) = B(m+1) = 0. \label{Eq31}
\end{equation}
This condition requires the following two equations to be satisfied
\begin{eqnarray}
gt\sqrt{C(m-1)}&=&k\pi, \label{Eq32}\\
gt\sqrt{C(m+1)}&=&gt\sqrt{C(m-1)}+l\pi=(l+k)\pi, \label{Eq33}
\end{eqnarray}
where $k$ and $l$ are both positive integers. The two equations above guarantee that both $B(m-1)$ and $B(m+1)$ are equal to zero, which means that both $w$ and $p$ are equal to 0. Also, when the above two equations hold, we can get $A(m-1)=\pm 1$ and $A(m+1)=\pm 1$. However, it is worth noting that $A(m+1)=1$ leads to $\mu=0$. Since our aim is to find the maximum entangled pure state of two qubits, we do not consider this case. Therefore, for the sake of discussion, we take
\begin{eqnarray}
A(m+1)&=&-1, \label{Eq34}\\
A(m-1)&=&1.  \label{Eq35}
\end{eqnarray}
We will discuss the case of $A(m+1)=A(m-1)=-1$ later. When Eq.~(\ref{Eq34}) and Eq.~(\ref{Eq35}) hold, through Eq.~(\ref{Eq33}), i.e., $gt\sqrt{C(m+1)}-gt\sqrt{C(m-1)}=\pi$, we can get the time when Eq.~(\ref{Eq34}) and Eq.~(\ref{Eq35}) first hold as follows
\begin{equation}
t_{1}=\frac{1}{g}\frac{\pi}{\sqrt{4m+6}-\sqrt{4m-2}}. \label{Eq36}
\end{equation}
Then, at the time $t_{1}$, we get the matrix elements of two qubits as follows
\begin{eqnarray}
v_{+}&=&\left|c_{m+2}\right|^{2}\frac{4m^{2}+12m+8}{4m^{2}+12m+9}, \label{Eq37}\\
w&=&p=0, \label{Eq38}\\
\mu&=&-c_{m}c_{m+2}^{*}\sqrt{\frac{4m^{2}+12m+8}{4m^{2}+12m+9}}, \label{Eq39}\\
v_{-}&=&1-\left|c_{m+2}\right|^{2}\frac{4m^{2}+12m+8}{4m^{2}+12m+9}. \label{Eq40}
\end{eqnarray}
Obviously, when we choose $c_{m}=1/\sqrt{2}$, $c_{m+2} = e^{-i(\phi+\pi)}/\sqrt{2}$, and $m$ tends to be large, then $v_{+}=v_{-}=1/2$, $\mu=e^{i\phi}/2$. This means that when we choose the initial state of the two identical qubits as $|g\rangle\otimes|g\rangle$ and the initial state of the optical field as a superposition of the following two number states
\begin{equation}
|\phi_{a}(0)\rangle=\frac{1}{\sqrt{2}}(|m\rangle+ e^{-i(\phi+\pi)}|m+2\rangle). \label{Eq41}
\end{equation}
If the value of $m$ is large enough to make $(4m^{2}+12m+8)/(4m^{2}+12m+9)=1$, then at time $t_{1}$, we can obtain the maximum entangled pure state of the two qubits as described by
\begin{eqnarray}
\hat{\rho}_{q}(t_{1})=\left[\begin{array}{cccc}
1/2 & 0 & 0 & e^{-i\phi}/2\\
0 & 0 & 0 & 0\\  
0 & 0 & 0 & 0\\
e^{i\phi}/2 & 0 & 0 & 1/2        
\end{array}\right].  \label{Eq42}
\end{eqnarray}

\begin{figure}[t]
\centering
\begin{tikzpicture}
\draw (0, 0) node[inner sep=0] {\includegraphics[width=8cm,height=5cm]{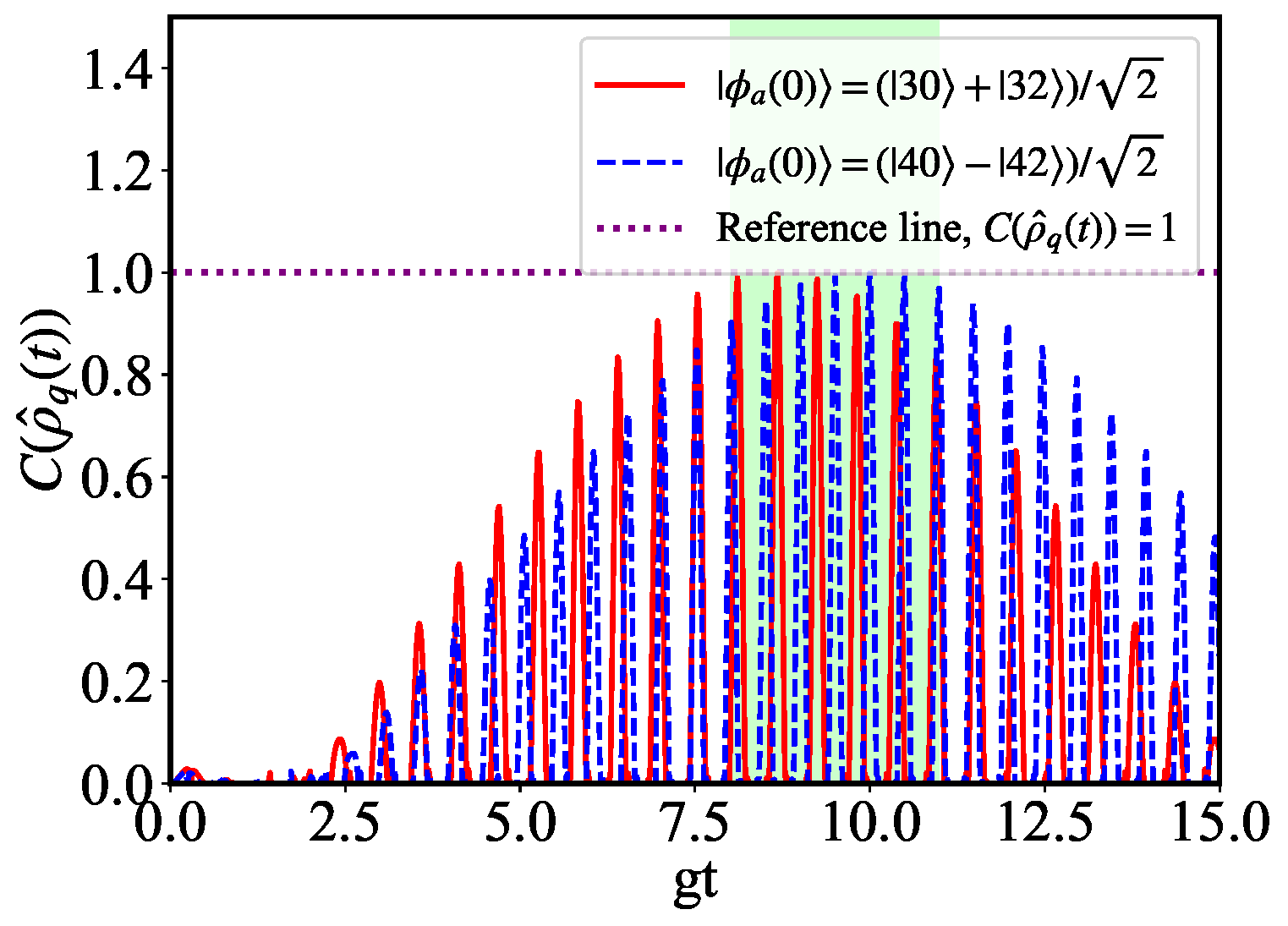}};
\draw (-2.3, 1.8) node {(a)};
\end{tikzpicture}
\begin{tikzpicture}
\draw (0, 0) node[inner sep=0] {\includegraphics[width=8cm,height=5cm]{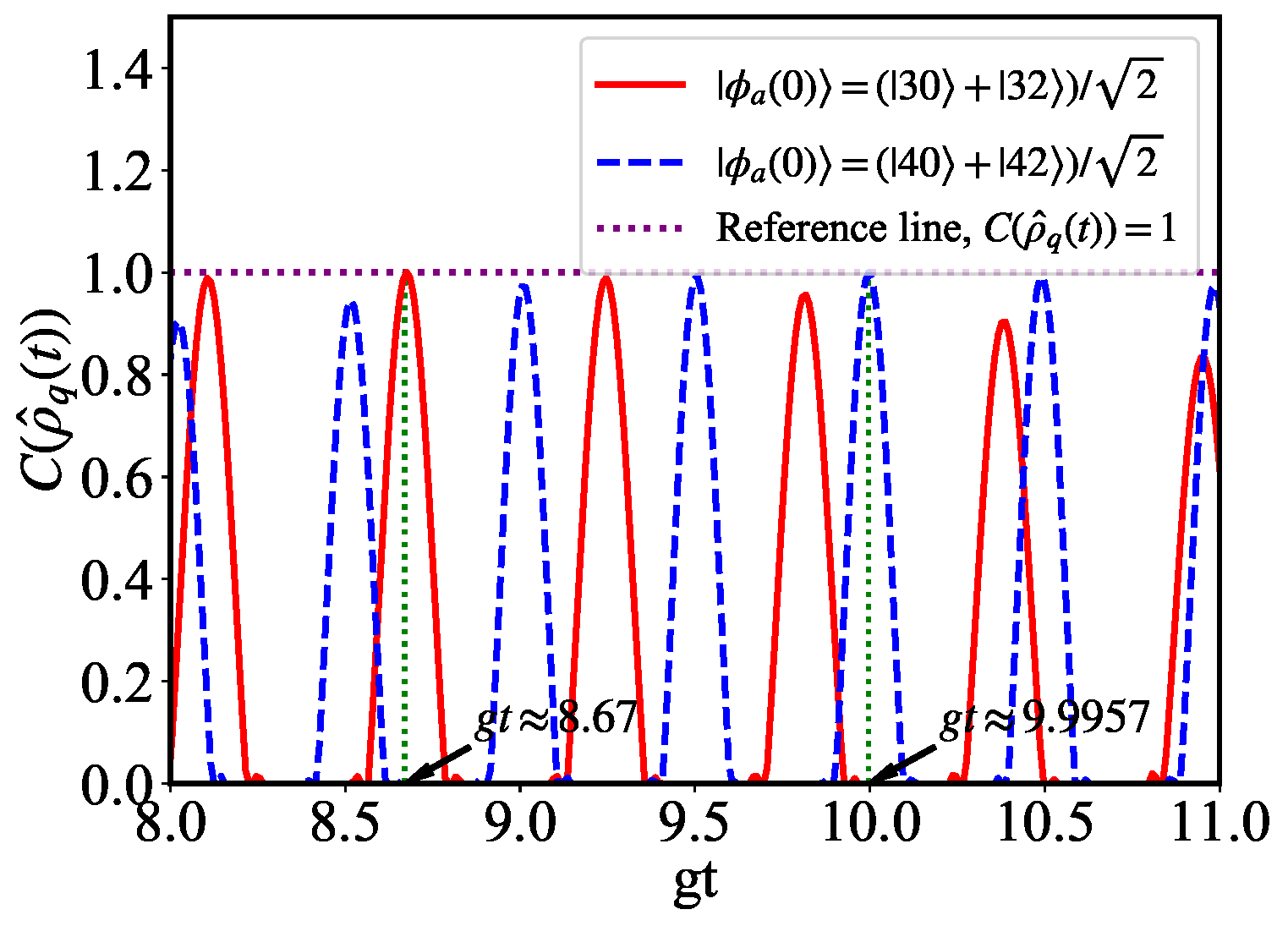}};
\draw (-2.3, 1.8) node {(b)};
\end{tikzpicture}
\caption{\label{fig3} (a) The red solid line and the blue dashed line indicate the change of the Concurrence of the two qubits with time when the initial state of the optical field is $|\phi_{a}(0)\rangle=(|30\rangle+ |32\rangle)/\sqrt{2}$ and $|\phi_{a}(0)\rangle=(|40\rangle- |42\rangle)/\sqrt{2}$, respectively. The purple dotted line is the reference line when the quantum entanglement takes its maximum value. (b) The image represents a zoomed-in view of the green region in the (a) figure.}
\end{figure}

In order to verify the above conclusions, we start to study the matrix elements of the density matrix of the two qubits system and the quantum entanglement between two qubits over time. For a quantum state $\hat{\rho}_{q}(t)$ of two qubits at time $t$, the magnitude of quantum entanglement is measured by the following function \cite{PhysRevLett.80.2245}
\begin{eqnarray}
E(C(\hat{\rho}_{q}(t)))&=&h(\frac{1+\sqrt{1-C^{2}(\hat{\rho}_{q}(t))}}{2}),  \label{Eq43}\\
h(x)&=&-x\log_{2}x-(1-x)\log_{2}(1-x), \label{Eq44}
\end{eqnarray}
where $E(C(\hat{\rho}_{q}(t)))$ is the entanglement of formation, and 
\begin{equation}
C(\hat{\rho}_{q}(t))=\max\{0, \lambda_{1}-\lambda_{2}-\lambda_{3}-\lambda_{4}\} \label{Eq45}
\end{equation}
is referred to as the quantum concurrence. $\lambda_{i}$ ($i=1,2,3,4$) is the square root of the eigenvalues of the matrix $\hat{\rho}_{q}(t)\hat{\varrho}_{q}(t)$ in descending order, that is,
$\lambda_{1}>\lambda_{2}>\lambda_{3}>\lambda_{4}$. $\hat{\varrho}_{q}(t)=(\hat{\sigma}_{y}\otimes\hat{\sigma}_{y})\hat{\rho}_{q}^{*}(t)(\hat{\sigma}_{y}\otimes\hat{\sigma}_{y})$, $\hat{\sigma}_{y}$ is the Pauli-Y operator, and $\hat{\rho}_{q}^{*}(t)$ is the complex conjugate of $\hat{\rho}_{q}(t)$. Since when the quantum concurrence $C(\hat{\rho}_{q}(t))$ monotonically changes from $0$ to $1$, the function $E(C(\hat{\rho}_{q}(t)))$ correspondingly monotonically changes from $0$ to $1$, quantum concurrence $C(\hat{\rho}_{q}(t))$ is often directly used to measure the magnitude of quantum entanglement between two qubits \cite{PhysRevLett.80.2245}. In this article, we directly use the quantum concurrence $C(\hat{\rho}_{q}(t))$ to measure the extent of quantum entanglement between two identical qubits.

Below, we choose the initial state of the two qubits as $|g\rangle\otimes|g\rangle$ and the initial state of the optical field as the following equal probability superposition state of the two number states
\begin{equation}
|\phi_{a}(0)\rangle=\frac{1}{\sqrt{2}}(|30\rangle+ |32\rangle). \label{Eq46}
\end{equation}
By substituting the parameters from Eq.~(\ref{Eq46}) into Eq.~(\ref{Eq27}) to Eq.~(\ref{Eq30}), we can obtain all the matrix elements of the density matrix of two qubits at time $t$. We then plot the evolution of these matrix elements and the quantum concurrence between the qubits over time, as shown in Fig.~\ref{fig2}. Fig.~\ref{fig2}a displays the initial density matrix elements of the two qubits. Fig.~\ref{fig2}b shows the matrix elements when the two qubits first reach the maximum entangled state, with the initial state of the optical field as described in Eq.~(\ref{Eq46}). In Fig.~\ref{fig2}c, the red solid line and blue dashed line represent the variation of quantum concurrence between the qubits over time for initial optical field states $|\phi_{a}(0)\rangle=(|30\rangle+ |32\rangle)/\sqrt{2}$ and $|\phi_{a}(0)\rangle=(|40\rangle+ |42\rangle)/\sqrt{2}$, respectively. Fig.~\ref{fig2}d is a magnified view of Fig.~\ref{fig2}c for the time interval $[8, 11]$. From Fig.~\ref{fig2}c and Fig.~\ref{fig2}d, we learn that the first time the quantum entanglement between the two qubits reaches its maximum is at times $gt\approx8.67$ and $gt\approx9.9957$ for optical field initial states $|\phi_{a}(0)\rangle=(|30\rangle+ |32\rangle)/\sqrt{2}$ and $|\phi_{a}(0)\rangle=(|40\rangle+ |42\rangle)/\sqrt{2}$, respectively. Substituting $m=30$ and $m=40$ into Eq.~(\ref{Eq36}) shows consistency with the actual evolution times. Additionally, it is observed that a smaller $m$ value leads to an earlier achievement of maximum quantum entanglement. Eq.~(\ref{Eq36}) indicates that a larger $m$ value, which brings the denominator closer to $0$, implies a longer time and vice versa. However, it is important to note that when $m$ is small, the condition $(4m^{2}+12m+8)/(4m^{2}+12m+9)=1$ might not be met, and the resulting state is not the maximum entangled state. Therefore, in practical experiments, the selection of $m$ should consider both the time constraint and the $(4m^{2}+12m+8)/(4m^{2}+12m+9)=1$ condition for preparing the maximum entangled state.

However, when we take $A(m-1)=A(m+1)=-1$, then we can obtain
\begin{eqnarray}
v_{+}&=&\left|c_{m}\right|^{2}\frac{4m^{2}-4m}{(2m-1)^{2}}+\left|c_{m+2}\right|^{2}\frac{4m^{2}+12m+8}{4m^{2}+12m+9} , \label{Eq47}\\
w&=&p=0,  \label{Eq48}\\
\mu&=&c_{m}c_{m+2}^{*}\frac{\sqrt{4m^{2}+12m+8}}{(2m+3)(2m-1)},  \label{Eq49}\\
v_{-}&=&\left|c_{m}\right|^{2}\frac{1}{(2m-1)^{2}}+\left|c_{m+2}\right|^{2}\frac{1}{(2m+3)^{2}}. \label{Eq50}
\end{eqnarray}
Since, in this case, the maximum entangled state cannot be obtained for arbitrary values of $m$, especially when $m$ takes a larger value that leads to $v_{+}=1$ and $\mu=v_{-}=0$. To produce the maximum entangled state of the two qubits in this scenario, it is required that $v_{+}=v_{-}=0.5$ is satisfied. In fact, when Eq.~(\ref{Eq47}) and Eq.~(\ref{Eq50}) are equal and both equal to $0.5$, four sets of solutions can be derived as follow
\begin{eqnarray}
\left|c_{m}\right|^{2}&=&-1.01631, m=-2.47645,  \nonumber\\
or \left|c_{m}\right|^{2}&=&0.144845, m=-0.832344,  \nonumber\\
or \left|c_{m}\right|^{2}&=&0.660974, m=-0.101541,  \nonumber\\
or \left|c_{m}\right|^{2}&=&1.72874, m=1.41034.  \nonumber
\end{eqnarray}
Unfortunately, these solutions do not satisfy the constraints that $\left|c_{m}\right|^{2} \geq 0$ and $m$ is a non-negative integer. So, we will not discuss this case in detail. 

In summary, if we want to prepare the two qubits initially in the state $|g\rangle\otimes|g\rangle$ into a maximally entangled state as shown in Eq.~(\ref{Eq25}), we just need to set the initial state of the optical field as shown in Eq.~(\ref{Eq41}) (of course $m$ in Eq.~(\ref{Eq41}) needs to be large enough to make the condition $(4m^{2}+12m+8)/(4m^{2}+12m+9)=1$ hold), and then we will get a maximally entangled state of two qubits at the time $gt=\pi/(\sqrt{4m+6}-\sqrt{4m-2})$ as shown in Eq.~(\ref{Eq25}).

Of course, we can also prepare the two qubits from the initial state $|e\rangle\otimes|e\rangle$ to the maximally entangled state as shown in Eq.~(\ref{Eq25}) by controlling the initial state of the optical field. This case requires an initial state of the cavity field similar to the above, which we will not continue to discuss below.

\section{Preparation of the second class of Bell state}

In the following, we prepare another maximum entangled state of two qubits by adjusting the initial state of the optical field, and this maximum entangled state is as follows
\begin{equation}
|B_{2}\rangle=(|eg\rangle+|ge\rangle)/\sqrt{2} . \label{Eq51}
\end{equation}
The density operator form of the second type of maximum entangled pure state for two qubits is as follows
\begin{eqnarray}
\hat{\rho}_{m_{2}}&=&|B_{2}\rangle\langle B_{2}| \nonumber \\
&=&\left[\begin{array}{cccc}
0 & 0 & 0 & 0\\
0 & 0.5 & 0.5 & 0\\
0 & 0.5 & 0.5 & 0\\
0 & 0 & 0 & 0
\end{array}\right]. \label{Eq52}
\end{eqnarray}

\begin{figure}[t]
\centering
\begin{tikzpicture}
\draw (0, 0) node[inner sep=0] {\includegraphics[width=8cm,height=5cm]{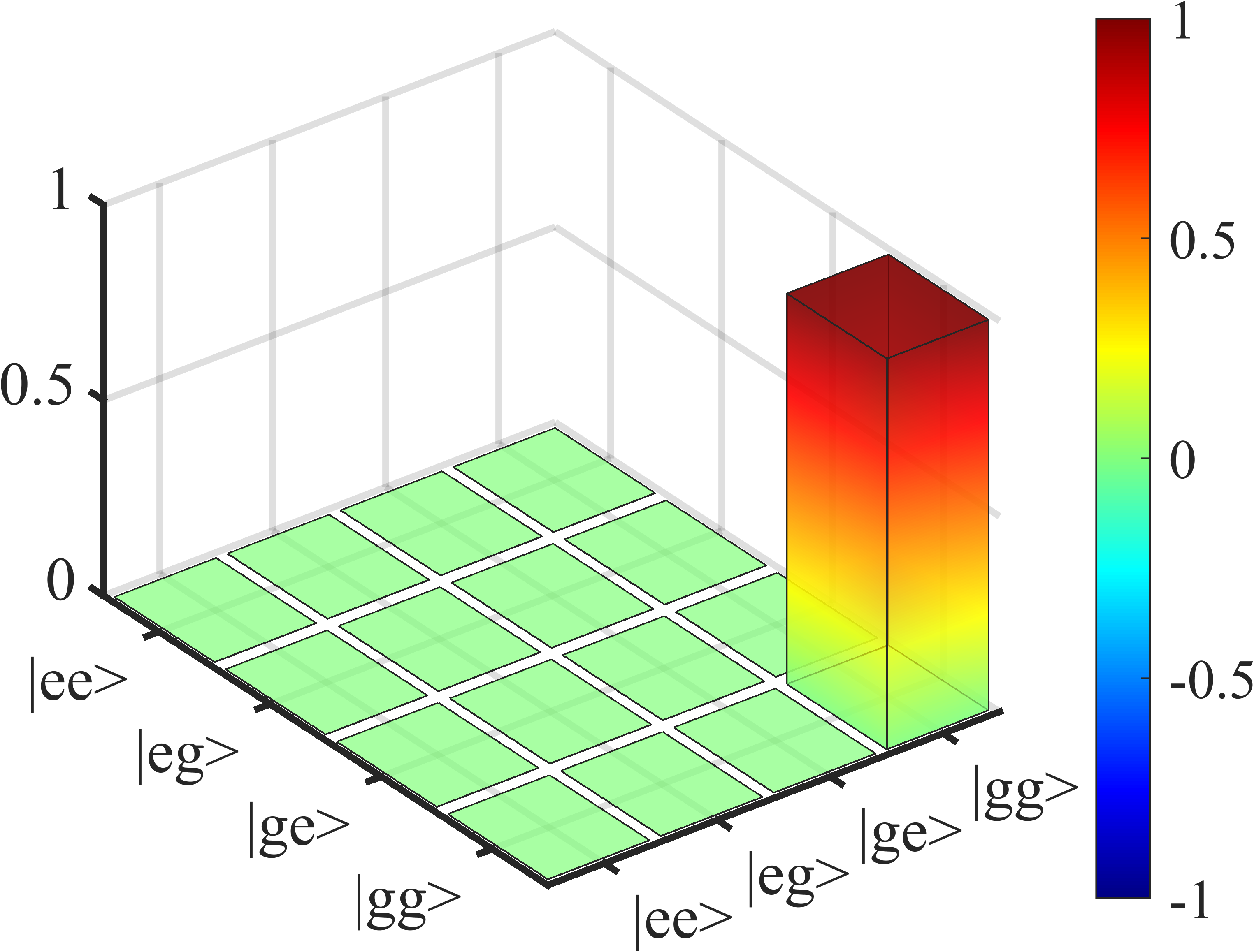}};
\draw (-2.5, 2) node {(a)};
\end{tikzpicture}
\begin{tikzpicture}
\draw (0, 0) node[inner sep=0] {\includegraphics[width=8cm,height=5cm]{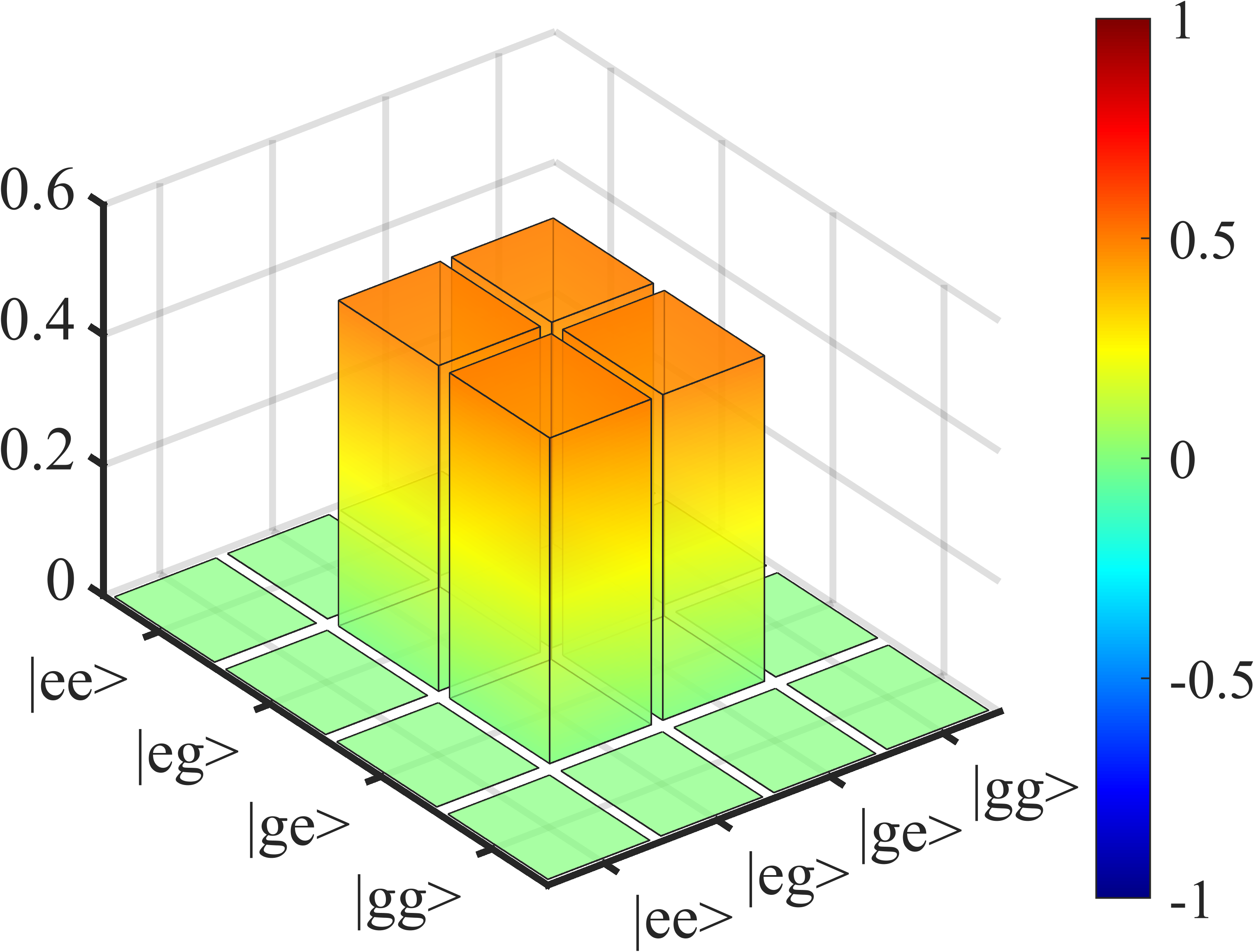}};
\draw (-2.5, 2) node {(b)};
\end{tikzpicture}
\caption{\label{fig4} When the initial state of the two qubits is $|gg\rangle$,  (a)  represents the image of the initial density matrix element of the two qubits, and  (b) represents the image of the matrix element of the density matrix of the two qubits when the Concurrence of the two qubits first reaches the maximum under the control of the optical field in the initial state of the single-photon number state $|m=1\rangle$.}
\end{figure}

Similar to the analysis above, if we aim to obtain a maximum entangled pure state as shown in Eq.~(\ref{Eq52}), then we can choose an initial number state of the optical field as follows
\begin{equation}
|\phi_{a}(0)\rangle=|m\rangle, \label{Eq53}
\end{equation}
where $m$ is a positive integer. When we choose such an optical field with a number state in the initial state, from Eq.~(\ref{Eq16}) to Eq.~(\ref{Eq21}) we get the matrix elements $h_{+}=h_{-}=\mu=0$ for the density matrix $\hat{\rho}_{1}(t)$ of two qubits at the moment $t$, and the other matrix elements are
\begin{equation}
w=p=\frac{m}{4m-2}\sin^{2}(gt\sqrt{2(2m-1)}).  \label{Eq54}
\end{equation}
Since $\textbf{Tr}(\hat{\rho}_{1}(t))=1$, both $v_{+}$ and $v_{-}$ are naturally equal to $0$ when $w =p= 0.5$, so we will not discuss the values of $v_{+}$ and $v_{-}$ here. In this case, we only need to consider the conditions when $w$ and $p$ are equal to $0.5$. But here is something worth noting, since $0\leq \sin^{2}(gt\sqrt{2(2m-1)}) \leq 1$, therefore $m/(4m-2)\geq0.5$. So $m\leq1$, and since $m$ is a positive integer, $m$ has only one value,  which is $m = 1$. Substituting $m=1$ into Eq.~(\ref{Eq54}), then
\begin{equation}
w=p=\frac{1}{2}\sin^{2}(gt\sqrt{2}). \label{Eq55}
\end{equation} 
When $w=p=0.5$, that is, the time when the state of the two qubits is a maximum entangled state shaped like Eq.~(\ref{Eq52}) is
\begin{equation}
t_{2}=\frac{l\pi}{2\sqrt{2}g}, \label{Eq56}
\end{equation}
where $l$ is an odd number. It implies that when we choose the following initial state of the optical field
\begin{equation}
|\phi_{a}(0)\rangle=|m=1\rangle, \label{Eq57}
\end{equation}
then two qubits can be prepared from the initial state $|g\rangle\otimes|g\rangle$ to the following maximally entangled pure state at the moment $t_{2}$
\begin{eqnarray}
\hat{\rho}_{q}(t_{2})=\left[\begin{array}{cccc}
0 & 0 & 0 & 0\\
0 & 0.5 & 0.5 & 0\\
0 & 0.5 & 0.5 & 0\\
0 & 0 & 0 & 0
\end{array}\right].  \label{Eq58}
\end{eqnarray}

\begin{figure}[t]
\centering
\includegraphics[width=8cm,height=5cm]{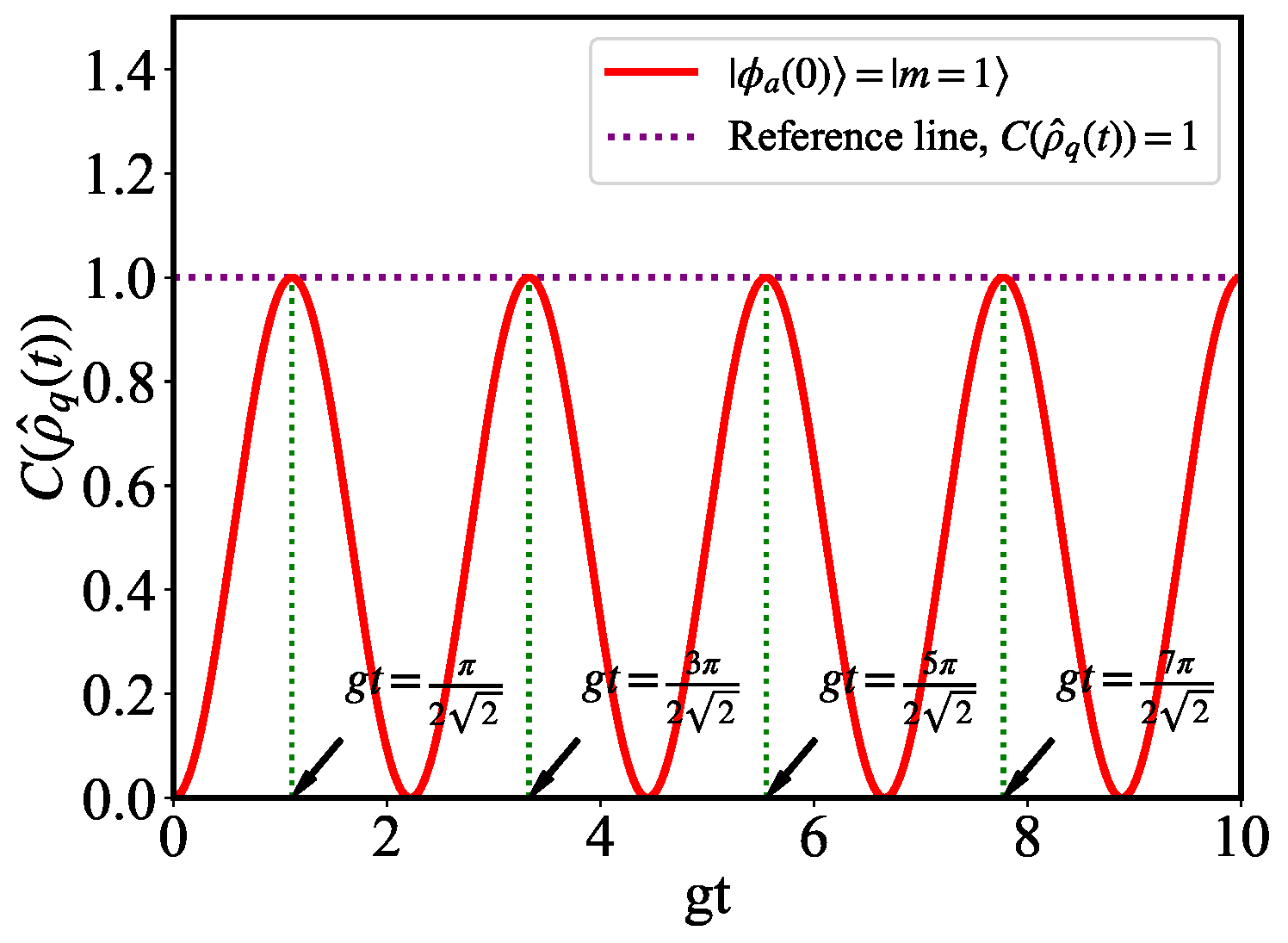}	
\caption{\label{fig5}Concurrence of two qubits with time under the control of an optical field with a single photon number state in the initial state when the initial state of the two qubits is $|gg\rangle$.}
\end{figure}

To validate the aforementioned conclusions, we begin to study the dynamic evolution of the system. The initial state of the optical field is chosen as a single photon number state $|m=1\rangle$, and the two qubits initially are in the product state $|g\rangle\otimes|g\rangle$ of two ground states. By substituting these initial conditions into Eq.~(\ref{Eq16}) to Eq.~(\ref{Eq21}), we can obtain all the matrix elements of the density matrix of the two qubits at any time. In Fig.~\ref{fig4}a and Fig.~\ref{fig4}b, we depict the matrix element images of the density matrix of the two qubits at the initial moment and when they reach the maximum entangled pure state, respectively. In Fig.~\ref{fig5}, we plot the change in the quantum concurrence of the two qubits over time, observing that the quantum entanglement between the two qubits varies periodically from none to maximum, with the maximum entanglement occurring at odd multiples of $\pi/(2\sqrt{2})$, consistent with the expected results from Eq.~(\ref{Eq55}). It is noteworthy that here we are unable to produce different types of maximum entangled pure states by adjusting the phase of a certain number state of the optical field, as is possible in the preparation of the first type of maximum entangled pure state. This result was already evident in Eq.~(\ref{Eq20}), where it is known that $w$ and $p$ are equal and can only be positive real numbers. Therefore, it is not possible to prepare maximum entangled states like the form of $(|eg\rangle-|ge\rangle)/\sqrt{2}$ here.

In summary, if we set the optical field initially in a single-photon number state $|m=1\rangle$ and both qubits are initially in the ground state $|g\rangle$, we can obtain a maximally entangled pure state of two qubits shaped like $|B_{2}\rangle=(|eg\rangle+|ge\rangle)/\sqrt{2}$ at the moment $gt=\frac{l\pi}{2\sqrt{2}}$, where $l$ is an odd number.

\section{Preparation of a two-qubits Werner state}

Werner states are a class of mixed quantum states that are commonly used in the study of quantum entanglement. These states were introduced by \textit{Werner} in 1989 \cite{PhysRevA.40.4277}. A Werner state for two qubits (two-level quantum systems) is defined as a statistical mixture of a maximally entangled state and a completely mixed state. Generally, a Werner state $\hat{\rho}_{w}$ is described as
\begin{equation}
\hat{\rho}_{w} = k |\Psi^{-}\rangle\langle\Psi^{-}| + (1 - k)\frac{\hat{I}}{4}. \label{Eq59}
\end{equation}
Here, $|\Psi^{-}\rangle=(|ge\rangle-|eg\rangle)/\sqrt{2}$ is the singlet state, which is a maximally entangled state of two qubits. $\hat{I}$ is a 4-dimensional identity matrix, and $\hat{I}/4$ represents the completely mixed state for two qubits. The parameter $k$ determines the mixture of these two states. When $k = 1$, $\hat{\rho}_{w}$ is a pure singlet state, and when $k = 0$, $\hat{\rho}_{w}$ is the completely mixed state. Werner states are particularly interesting because they can describe a range of entanglement properties. For $k < \frac{1}{3}$, the state is separable, meaning there is no entanglement. For $\frac{1}{3} < k < 1$, the state is partially entangled, and for $k = 1$, it is maximally entangled. These states are valuable in quantum information theory and quantum optics for studying the behavior of entanglement under different conditions and for exploring the boundaries between classical and quantum correlations. They serve as a useful tool in understanding quantum entanglement and its applications in quantum communication and quantum computing. Also, $\hat{\rho}_{w}$ can be written as
\begin{eqnarray}
\hat{\rho}_{w}=\left[\begin{array}{cccc}
\eta/3 & 0 & 0 & 0\\
0 & (3-2\eta)/6 & (-3+4\eta)/6 & 0\\
0 & (-3+4\eta)/6 & (3-2\eta)/6 & 0\\
0 & 0 & 0 & \eta/3
\end{array}\right] , \label{Eq60}
\end{eqnarray}
where $\eta=(3-3k)/4$.

\begin{figure}[t]
\centering
\includegraphics[width=8cm,height=5cm]{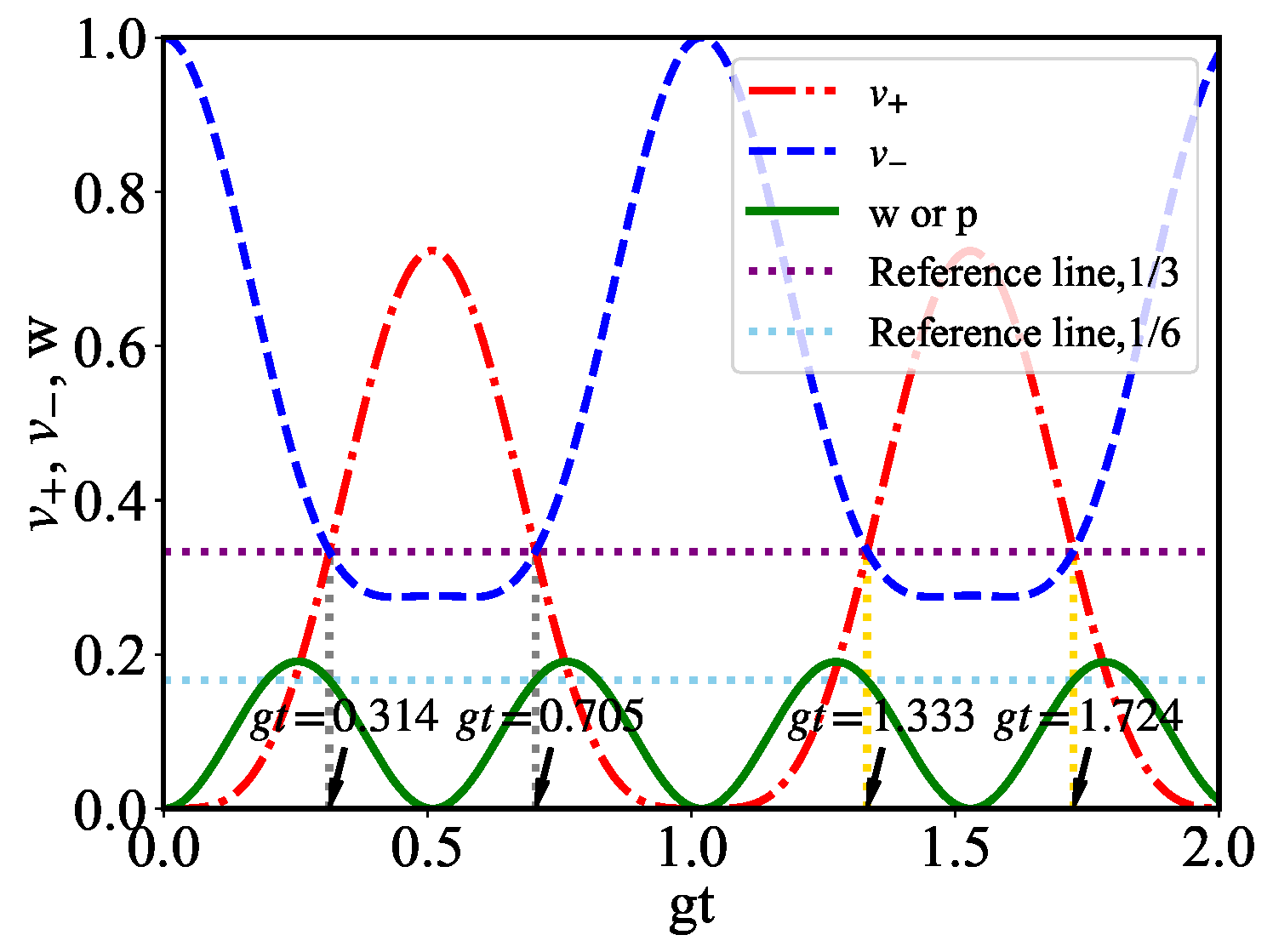}	
\caption{\label{fig6} When the initial state of the two qubits is $|gg\rangle$, the red, blue, and green dotted lines indicate the changes with time of the matrix elements $v_{+}$, $v_{-}$, and $w$, respectively, of the density matrix of the two qubits under the control of the optical field in the initial state $|\phi_{a}(0)\rangle=c_{0}|0\rangle+c_{10}|10\rangle$. The purple and sky blue dotted lines indicate the reference lines for matrix elements with values of $1/3$ and $1/6$, respectively.}
\end{figure}

In the following, we begin to study the preparation of two qubits from the initial state $|gg\rangle$ to a Werner state shaped as shown in Eq.~(\ref{Eq59}) by means of an optical field initial state. From Eq.~(\ref{Eq16})-Eq.~(\ref{Eq21}), we know that the matrix elements $w$ and $p$ of the two qubits are equal, i.e., for the Werner state shown in Eq.~(\ref{Eq59}), $(3-2\eta)/6=(-3+4\eta)/6$, then $\eta=1$. Thus, we are able to obtain just one Werner state, as shown in the following equation
\begin{eqnarray}
\hat{\rho}_{w}=\left[\begin{array}{cccc}
1/3 & 0 & 0 & 0\\
0 & 1/6 & 1/6 & 0\\
0 & 1/6 & 1/6 & 0\\
0 & 0 & 0 & 1/3
\end{array}\right].  \label{Eq61}
\end{eqnarray}

\begin{figure}[t]
\centering
\begin{tikzpicture}
\draw (0, 0) node[inner sep=0] {\includegraphics[width=8cm,height=5cm]{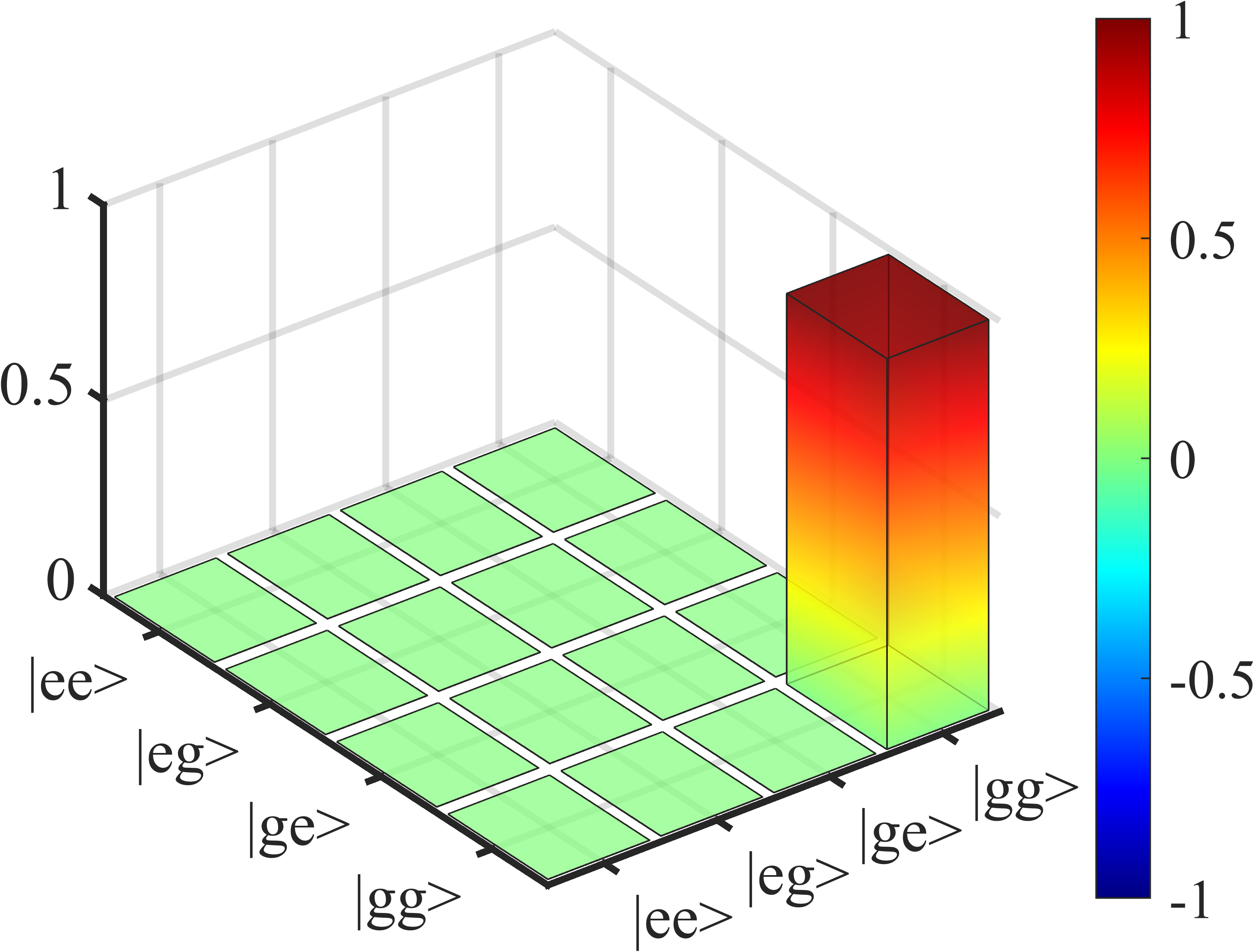}};
\draw (-2.5, 2) node {(a)};
\end{tikzpicture}
\begin{tikzpicture}
\draw (0, 0) node[inner sep=0] {\includegraphics[width=8cm,height=5cm]{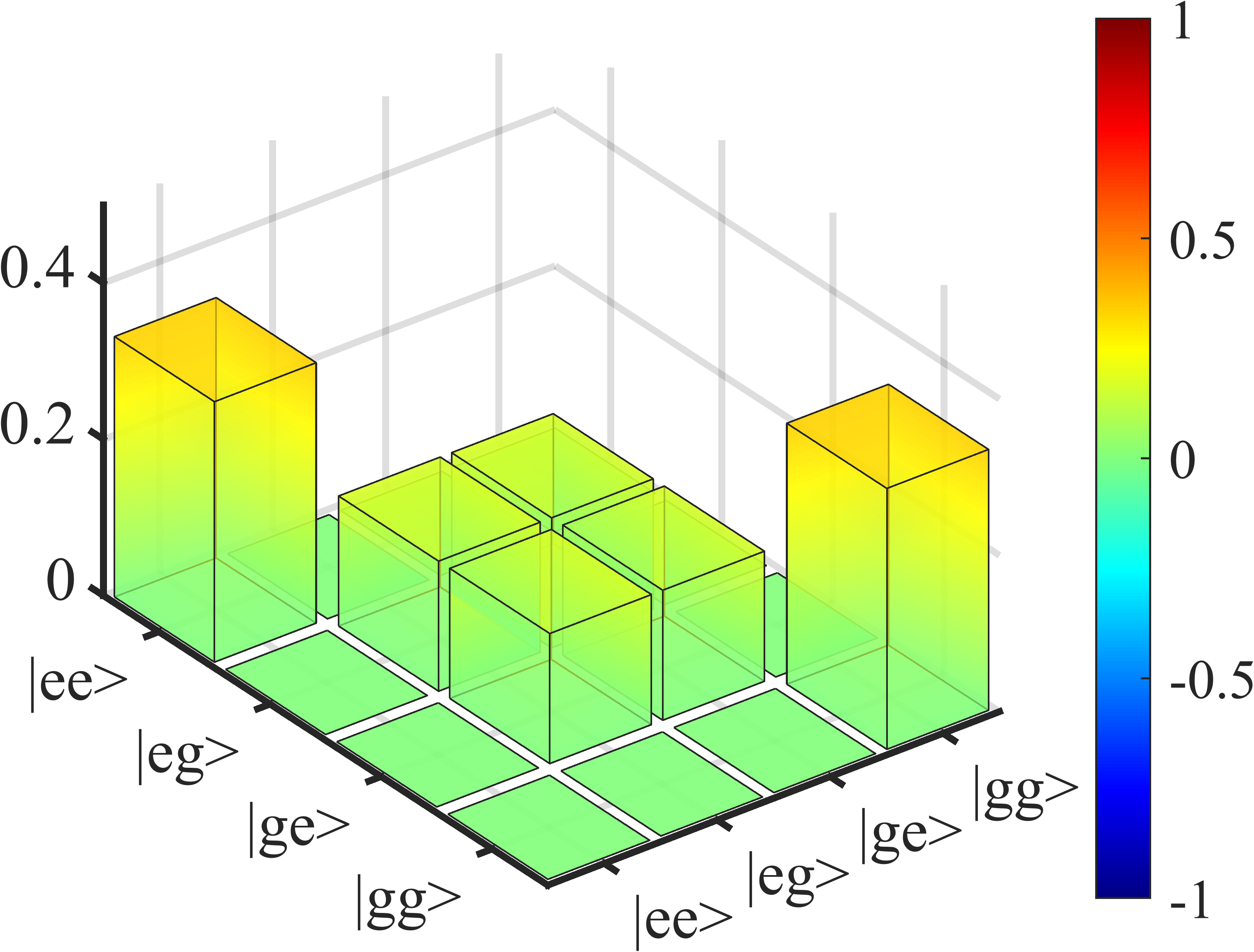}};
\draw (-2.5, 2) node {(b)};
\end{tikzpicture}
\caption{\label{fig7} When the initial state of the two qubits is $|gg\rangle$, (a) denotes the image of the matrix element of the density matrix of the two qubits, and (b) denotes the image of the matrix element of the density matrix of the two qubits when the two qubits evolve for the first time to the Werner state as shown in Eq.~(\ref{Eq60}) under the control of the optical field with an initial state of $|\phi_{a}(0)\rangle=c_{0}|0\rangle+c_{10}|10\rangle$.}
\end{figure}

From Eq.~(\ref{Eq16})-Eq.~(\ref{Eq21}), we know that $h_{+}=h_{-}=\mu=0$ if we choose an initial state of the optical field as follows
\begin{equation}
|\phi_{a}(0)\rangle=c_{0}|0\rangle+c_{10}|10\rangle , \label{Eq62}
\end{equation}
where $|c_{0}|^{2}+|c_{10}|^{2}=1$, and the other matrix elements of $\hat{\rho}_{1}(t)$
\begin{eqnarray}
v_{+}&=&\frac{90}{361}|c_{10}|^{2} ( \cos{\sqrt{38}gt}-1)^{2},  \label{Eq63}\\
v_{-}&=&|c_{0}|^{2} + |c_{10}|^{2} [1 + \frac{10}{19} (\cos{\sqrt{38}gt}-1)]^2 , \label{Eq64}\\
w&=&p=|c_{10}|^{2}\frac{5}{19}\sin^{2}{\sqrt{38}gt} \label{Eq65}.
\end{eqnarray}
From the system of equations $\{v_{+}=v_{-}=1/3, w=p=1/6\}$, we get the following solutions
\begin{eqnarray}
|c_{0}|^{2}&=&0.274,    \label{Eq66}\\
|c_{10}|^{2}&=&0.726   \label{Eq67}\\
gt&\approx&1.019l\pm0.314 , \label{Eq68}
\end{eqnarray}
where $l$ is an integer.

In Fig.~\ref{fig6}, when the initial state of the total system is $|\Psi(0)\rangle=|g\rangle\otimes|g\rangle\otimes(c_{0}|0\rangle+c_{10}|10\rangle)$, we plot the nonzero matrix elements of the two qubits separately over time $t$. The red, blue, and green dotted lines denote the matrix elements $v_{+}$, $v_{-}$, and $w$ over time $t$, respectively. The purple and sky blue dotted lines indicate the reference lines when the values of the matrix elements are equal to $1/3$ and $1/6$, respectively. Yellow dashed lines indicate the reference lines when $gt$ takes different values. From the figure, we can see that, in agreement with the expected time through Eq.~(\ref{Eq68}), the matrix elements $v_{+}$, $v_{-}$, and $w$ take values of $1/3$, $1/3$, and $1/6$, respectively, when $gt = 1.019l\pm0.314$. In order to validate the above results, we have plotted the images of the matrix elements of the two qubits at the initial moment and at the time of the first attainment of the Werner state in Figs.~\ref{fig7}a and ~\ref{fig7}b, respectively. From Fig.~\ref{fig7}b, we can see that the density matrix of the two qubits at this time is a Werner state, as shown in Eq.~(\ref{Eq61}).

In summary, it is shown that when the initial state of the two qubits is $|gg\rangle$, we can control the two qubits to produce a Werner state, as shown in Eq.~(\ref{Eq61}), by choosing the optical field whose initial state is $|\phi_{a}(0)\rangle=c_{0}|0\rangle+c_{10}|10\rangle$, where the values of $c_{0}$ and $c_{10}$ are shown in Eq.~(\ref{Eq65}) and Eq.~(\ref{Eq66}), respectively.

\section{\label{Sec:4} Conclusion }
We find that controlling the optical field in different initial states can control the two qubits initially in the direct product state to produce the first type of Bell state, the second type of Bell state, and the Werner state, respectively. By solving and analyzing the two qubits TC model, we find that the matrix elements of the two qubits at any moment can be controlled by the population coefficients of the initial states of the optical field in the number state basis. Firstly, we found that as long as the expansion of the optical field state in the number state basis does not involve neighboring number states, it is possible to control the generation of an \textit{X}-type state in the two qubits. Second, when both qubits are initially in the ground state, an optical field that is initially in the next-nearest-neighbor number state is able to control the two qubits to produce any of the first type Bell states. Notably, the average photon number of such number state superposition states needs to be sufficiently large. Then, when one of the two qubits is initially in the ground state, and the other is initially in the excited state, the optical field that is initially in the single-photon number state is able to control the two qubits to produce a second type of Bell state. It is worth noting that the second type of Bell state produced at this point is not arbitrary. Finally, we find that the two qubits, both of which are initially in the ground state, can be controlled to produce Werner states by the superposition of the two farther-adjacent number states.

\begin{acknowledgments}
This work was supported by the NSFC (Grants No. 12205092, No. 12381240349, and No. 12205088), the Scientific Research Fund of Hunan Provincial Education Department (Grants
No. 22A0507, No. 21B0647, No. 21B0639), and the Open fund project of the Key
Laboratory of Optoelectronic Control and Detection Technology of University of Hunan Province
(Grant No. 2022HSKFJJ038)
\end{acknowledgments}  

\appendix

\section{\label{Appendix A}Detailed calculations for Eq.~(\ref{Eq16})-Eq.~(\ref{Eq21})}

In this appendix, we give the detailed calculations for Eq.~(\ref{Eq16})-Eq.~(\ref{Eq21}). In the following calculations, we have used the formulas $\hat{a}^{\dagger}f(\hat{n})=f(\hat{n}-1)\hat{a}^{\dagger}$ and $\hat{a}f(\hat{n})=f(\hat{n}+1)\hat{a}$ \cite{louisell1973quantum}.

\begin{eqnarray}
v_{+}&=&\sum_{n=0}^{\infty}\langle n|\hat{U}_{14}|\phi_{a}(0)\rangle\langle\phi_{a}(0)|\hat{U}_{14}^{\dagger}|n\rangle  \nonumber\\
&=&\sum_{n=0}^{\infty}\langle n|\hat{U}_{14}\text{\ensuremath{\sum_{p=0}^{\infty}|p\rangle\langle p|}}\phi_{a}(0)\rangle\langle\phi_{a}(0)|\sum_{q=0}^{\infty}|q\rangle\langle q|\hat{U}_{14}^{\dagger}|n\rangle   \nonumber\\
&=&\sum_{n=0}^{\infty}\left|c_{n+2}\right|^{2}4(n+2)(n+1)\left[\frac{A(n+1)-1}{C(n+1)}\right]^{2}, \\
h_{+}^{*}&=&\sum_{n=0}^{\infty}\langle n|\hat{U}_{14}|\phi_{a}(0)\rangle\langle\phi_{a}(0)|\hat{U}_{24}^{\dagger}|n\rangle  \nonumber\\
&=&\sum_{n=0}^{\infty}\langle n|\hat{U}_{14}\sum_{p=0}^{\infty}|p\rangle\langle p|\phi_{a}(0)\rangle\langle\phi_{a}(0)|\sum_{q=0}^{\infty}|q\rangle\langle q|\hat{U}_{24}^{\dagger}|n\rangle  \nonumber\\
&=&\sum_{n=0}^{\infty}c_{n+1}^{*}c_{n+2}2\sqrt{(n+2)(n+1)}\frac{A(n+1)-1}{C(n+1)}(i \nonumber \\
&&\times\sqrt{n+1}\frac{B(n)}{\sqrt{C(n)}})   \nonumber\\
&=&\sum_{n=0}^{\infty}c_{n+1}^{*}c_{n+2}(2i(n+1))\sqrt{(n+2)}\frac{A(n+1)-1}{C(n+1)} \nonumber \\
&&\times\frac{B(n)}{\sqrt{C(n)}} ,   \\
h_{-}^{*}&=&\sum_{n=0}^{\infty}\langle n|\hat{U}_{24}|\phi_{a}(0)\rangle\langle\phi_{a}(0)|\hat{U}_{44}^{\dagger}|n\rangle  \nonumber\\
&=&\sum_{n=0}^{\infty}\langle n|\hat{U}_{24}\sum_{p=0}^{\infty}|p\rangle\langle p|\phi_{a}(0)\rangle\langle\phi_{a}(0)|\sum_{q=0}^{\infty}|q\rangle\langle q|\hat{U}_{44}^{\dagger}|n\rangle    \nonumber  \\
&=&\sum_{n=0}^{\infty}c_{n}^{*}c_{n+1}(-i\sqrt{n+1}\frac{B(n)}{\sqrt{C(n)}})\Big(2\frac{A(n-1)-1}{C(n-1)}n \nonumber \\
&&+1\Big) , \\
\mu^{*}&=&\sum_{n=0}^{\infty}\langle n|\hat{U}_{14}|\phi_{a}(0)\rangle\langle\phi_{a}(0)|\hat{U}_{44}^{\dagger}|n\rangle  \nonumber\\
&=&\sum_{n=0}^{\infty}\langle n|\hat{U}_{14}\sum_{p=0}^{\infty}|p\rangle\langle p|\phi_{a}(0)\rangle\langle\phi_{a}(0)|\sum_{q=0}^{\infty}|q\rangle\langle q|\hat{U}_{44}^{\dagger}|n\rangle    \nonumber\\
&=&\sum_{n=0}^{\infty}c_{n}^{*}c_{n+2}2\sqrt{(n+2)(n+1)}\Big(1+2\frac{A(n-1)-1}{C(n-1)}n\Big) \nonumber \\
&&\times\frac{A(n+1)-1}{C(n+1)},  \\
w&=&p  \nonumber\\
&=&\sum_{n=0}^{\infty}\langle n|\hat{U}_{24}|\phi_{a}(0)\rangle\langle\phi_{a}(0)|\hat{U}_{24}^{\dagger}|n\rangle  \nonumber\\
&=&\sum_{n=0}^{\infty}\langle n|\hat{U}_{24}\sum_{p=0}^{\infty}|p\rangle\langle p|\phi_{a}(0)\rangle\langle\phi_{a}(0)|\sum_{q=0}^{\infty}|q\rangle\langle q|\hat{U}_{24}^{\dagger}|n\rangle  \nonumber\\
&=&\sum_{n=0}^{\infty}\left|c_{n+1}\right|^{2}(n+1)\frac{B^{2}(n)}{C(n)}, \\
v_{-}&=&\sum_{n=0}^{\infty}\langle n|\hat{U}_{44}|\phi_{a}(0)\rangle\langle\phi_{a}(0)|\hat{U}_{44}^{\dagger}|n\rangle  \nonumber\\
&=&\sum_{n=0}^{\infty}\langle n|\hat{U}_{44}\sum_{p=0}^{\infty}|p\rangle\langle p|\phi_{a}(0)\rangle\langle\phi_{a}(0)|\sum_{q=0}^{\infty}|q\rangle\langle q|\hat{U}_{44}^{\dagger}|n\rangle  \nonumber\\
&=&\sum_{n=0}^{\infty}\left|c_{n}\right|^{2}\left(1+2\frac{A(n-1)-1}{C(n-1)}n\right)^{2}.
\end{eqnarray}
where $c_{n}=\langle n|\phi_{a}(0)\rangle$, $|n\rangle$ and $\hat{I}_{c}=\sum_{p=0}^{\infty}|p\rangle\langle p|=\sum_{q=0}^{\infty}|q\rangle\langle q|$ are the number state and identity operator of the optical field, respectively.

\nocite{*}

\bibliography{PMESXS}% Produces the bibliography via BibTeX.

\end{document}